\documentclass[reprint,showpacs,twocolumn,superscriptaddress,aps,prx]{revtex4-2}
\usepackage[utf8]{inputenc}
\usepackage[american,british]{babel}
\usepackage[T1]{fontenc}
\usepackage[pdftex]{graphicx}  
\usepackage{graphicx, xcolor}
\usepackage{dcolumn}
\usepackage{physics}
\usepackage{braket}
\usepackage{bm}
\usepackage{amsmath,amsthm,amssymb}
\usepackage{color}
\usepackage{verbatim} 
\usepackage{soul}

\usepackage{ulem}
\usepackage{natbib}

\newcommand{\lptms}{Universit\'e Paris-Saclay, CNRS, LPTMS, 91405, Orsay, France}
\newcommand{\osnabruck}{Department of Mathematics/Computer Science/Physics, University of Osnabr\"uck, D-49076 Osnabr\"uck, Germany} 
\newcommand{\smt}{Science, Mathematics and Technology Cluster, Singapore University of Technology and Design, 8 Somapah Road, 487372 Singapore}
\newcommand{\epd}{Engineering Product Development Pillar, Singapore University of Technology and Design, 8 Somapah Road, 487372 Singapore}
\newcommand{\cqt}{Centre for Quantum Technologies, National University of Singapore 117543, Singapore}
\newcommand{\majulab}{MajuLab, CNRS-UNS-NUS-NTU International Joint Research Unit, UMI 3654, Singapore} 
\newcommand{\sichuan}{College of Physics and Electronic Engineering, and Center for Computational Sciences, Sichuan Normal University, Chengdu 610068, China} 
    
\definecolor{darkGreen}{RGB}{0,110,0}
\definecolor{darkBlue}{RGB}{0,0,130}
\usepackage{hyperref}

\hypersetup{
 colorlinks=true,
 linkcolor=blue,
 anchorcolor = blue,
 citecolor = blue,
 filecolor = blue,
 urlcolor = blue
}

\usepackage{dsfont}

\def \be {\begin{equation}} 
\def \ee {\end{equation}} 
\def \l {\left(} 
\def \r {\right)} 
\def \la {\langle} 
\def \ra {\rangle}  

\usepackage{tabularx}

\def\normOrd#1{\mathop{:}\nolimits\!#1\!\mathop{:}\nolimits}

\begin{document}

\title{Hydrodynamics and the eigenstate thermalization hypothesis}

\author{Luca Capizzi} 
\thanks{These authors contributed equally.}
\affiliation{\lptms}

\author{Jiaozi Wang} 
\thanks{These authors contributed equally.} 
\affiliation{\osnabruck} 

\author{Xiansong Xu} 
\affiliation{\smt} 
\affiliation{\sichuan}

\author{Leonardo Mazza}
\email{leonardo.mazza@universite-paris-saclay.fr}
\affiliation{\lptms}

\author{Dario Poletti}
\email{dario\_poletti@sutd.edu.sg}
\affiliation{\smt} 
\affiliation{\epd} 
\affiliation{\cqt} 
\affiliation{\majulab}

\begin{abstract}
The eigenstate thermalization hypothesis (ETH) 
describes the properties of diagonal and off-diagonal matrix elements of local operators in the eigenenergy basis. 
In this work, we propose a relation between (i) the singular behaviour of the off-diagonal part of ETH at small energy differences, and (ii) the smooth profile of the diagonal part of ETH as a function of the energy density. We establish this connection from the decay of the autocorrelation functions of local operators, which is constrained by the presence of local conserved quantities whose evolution is described by hydrodynamics. We corroborate our predictions through numerical simulations of two distinct non-integrable spin-1 Ising models, exhibiting diffusive and super-diffusive transport behaviors. The simulations are performed using dynamical quantum typicality up to 18 spins, for both infinite and finite temperature regimes.
\end{abstract}

\maketitle

\section{Introduction}

One of the deepest questions in statistical mechanics concerns the dynamical emergence of thermal equilibrium within isolated many-particle systems \cite{Bryan-02,vNeumann-29,Landau-80}.
In the classical Hamiltonian context, answering these questions has proven one of the most fruitful exercises in mathematical physics, with the development of paradigmatic notions, such as that of chaos, or results, such as the H-theorem, that shape our current description of physical phenomena \cite{Boltzmann-70,Boltzmann-70a,Arnold-09,Arnold-63,Huang-08}. 
It comes as no surprise that the history of the answers given to this question in the context of quantum mechanics is similarly disseminated of beautiful ideas and insightful intuitions~\cite{gltz-06,psw-06,pssv-11,dkpr-16, Essler_2016, Vidmar_2016}.

One remarkable framework for describing the emergence of thermalization from quantum dynamics in an isolated quantum many-body system is to postulate the eigenstate thermalization hypothesis (ETH). The current formulation of ETH, used in this article, is due to M.~Srednicki~\cite{Srednicki-99}, close to previous works by J.M.~Deutsch~\cite{Deutsch-91} and M.~Berry~\cite{berry1977}. Without entering into the fine-tuned details of the specific microscopic models under discussion, ETH proposes a view on its energy eigenstates as random vectors characterised by a unique quantum number, the energy density.
This simple postulate is sufficient to derive the dynamical emergence of a thermodynamic behaviour, reconciling the reversible Schr\"odinger equation with the directional arrow of time typical of macroscopic phenomena~\cite{pssv-11,dkpr-16}.

In practice, according to ETH, the matrix elements of a local observable $\mathcal O$ between two energy eigenstates $\ket{E_i}$ and $\ket{E_j}$ is
\begin{equation}\label{eq:ETH}
   \bra{E_i}\mathcal{O}\ket{E_j} = \mathcal O(\varepsilon) \delta_{ij} + e^{- \frac{S(\varepsilon)}{2}} 
   f_{\mathcal O}(\varepsilon, \omega) R_{ij}.
\end{equation}
Here, $\varepsilon$ is the average energy density $\varepsilon = (E_i + E_j)/2L$, with $L$ the system size, and the energy difference is $\omega = E_i - E_j$. The diagonal part of ETH postulates the existence of a smooth function $\mathcal O(\varepsilon)$ that describes the expectation value of $\mathcal O$ as a function of the energy density $\varepsilon = E_i/L$. 

This fact in itself has already important implications, as we can see with the following simple example.
An initial uncorrelated quantum many-body state $\ket{\Psi}$ is evolved unitarily in time with the Hamiltonian $H$ and our focus is on the expectation value as a function of time of a local operator, $\langle \mathcal O \rangle_t$.
Since the initial product state has a given energy density $\varepsilon_\Psi$ and sub-extensive energy fluctuations, we can conclude that at late times, when all eigenstates have dephased, the expectation value $\langle \mathcal O \rangle_t $ will coincide with $\mathcal O(\varepsilon_\Psi)$.
The relation with thermodynamics is given by the fact that the latter corresponds to the microcanonical expectation value of the observable.

Remarkably, ETH can also be used to derive predictions concerning late but finite times, describing the final transient dynamics before reaching the stationary thermodynamic values.
This is possible because the off-diagonal part of ETH proposes a form also for the matrix elements of the operator $\mathcal O$ between different energy eigenstates with average energy density $\varepsilon$.
The leading contribution depends on the number of such eigenstates, scaling as $e^{+ S(\varepsilon)}$  according to the Boltzmann entropy definition at energy density $\varepsilon$; this explains the prefactor $e^{- \frac{S(\varepsilon)}{2}}$ in Eq.~\eqref{eq:ETH}.
Moreover, the function $f_{\mathcal O}(\varepsilon, \omega)$ describes, together with the random variable $R_{ij}$ with zero average and unit variance, a fine structure as a function of the energy difference $\omega$.
The statistical properties of the random variables $R_{ij}$ have been deeply characterized in Refs.~\cite{Prosen-94,Kurchan19-GETH, GETH-Foini19, lmvr-19,rdsg-20,Wang22-ETH-offdiag, Dymarsky22-ETH-offdiag, pfk-22,Wang23-ETH-offdiag, GETH-Silvia}.
In the spirit of the above example, if instead of studying the local operator $\langle \mathcal O \rangle_t$ we rather focus on the connected autocorrelation function $\langle \mathcal O(t) \mathcal O(0) \rangle_{c}$, it is not unexpected that its late-time limit $t \to \infty$ is related to the low-frequency limit $\omega \to 0$ of $f_{\mathcal O}(\varepsilon_\Psi, \omega )$.

In summary, ETH is a conceptually simple and physically-motivated postulate that is capable of describing some of the most crucial properties of the dynamics of a quantum many-body system with limited assumptions, among which the existence of the functions $\mathcal O(\varepsilon)$ and $f_{\mathcal O}(\varepsilon, \omega)$. 
In recent years, extensive numerical experiments in lattice systems have been performed, providing compelling evidence of eigenstate thermalization for diagonal and off-diagonal matrix elements of local observables in a wide variety of models~\cite{rdo-08,Rigol-09,sr-10,rs-10,bkl-10,krrg-12,ghl-12,bmh-14,bmh-15,kpsr-13,shp-13,kih-14,skngg-14,ksg-15,mfsr-16}.

Despite this success, it is crucial to remember that ETH represents just one of several approaches to quantitatively describing the late-time dynamics of quantum many-body systems.
For example, the autocorrelation functions that we just discussed are a specific instance of the correlation functions that have been recently placed under the heavy scrutiny of hydrodynamic theories~\cite{km-63,bkv-05,baw-06,msf-09,Spohn-12,lmmr-14,ngtsm-15,ttgp-15,mknsm-16,Crossno-16,lf-18,cdy-16,bcdf-16,dbd-19,Doyon-22}.
This is a completely different framework that proposes a coarse-grained view of the dynamics via the introduction of mesoscopic fluid cells and focuses on the universal spreading of conserved quantities through the quantum many-body system.
Under the general assumption that energy, momentum and particles spread ballistically, diffusively or anomalously, hydrodynamics is a powerful tool for predicting the long-time behaviour of several properties, among which generic autocorrelation functions. 

The two aforementioned approaches to the long-time dynamics of quantum many-body systems have never been put in connection, even if one could easily anticipate that such an analysis is of great interest. Connecting the properties of the ETH functions to the spreading of the energy density could for instance ease the numerical inspection of the dynamics, or allow for the development of novel tools for computing linear and non-linear responses on top of equilibrium states. Viceversa, it could be extremely exciting to ascertain whether the knowledge derived from hydrodynamics that energy spreads diffusively or ballistically puts any constraint on the aforementioned ETH functions.

The goal of the article is twofold.
First, we want to revisit ETH and show that by studying the behaviour of autocorrelators one can link the properties of the two ETH functions, $\mathcal O (\varepsilon)$ and $f_{\mathcal O}(\varepsilon, \omega)$.
So far, these two functions have been always considered as completely independent and unrelated: here we show that this is not the case.
In particular, by inspecting the singular behaviour of $f_{\mathcal O}(\varepsilon, \omega)$ in the limit $\omega \to 0$, we can place bounds on the fact that a number of derivatives $\frac{d^m }{d \varepsilon^m} \mathcal O(\varepsilon)$ should equal zero using what we refer to as the relaxation-overlap inequality.
In order to do that we need to postulate the type of spreading of the energy density, invoking a hydrodynamics viewpoint on the quantum many-body dynamics. The direct physical consequence of the aforementioned singularities of $\omega$ is the presence of slow algebraic decaying in time of the autocorrelator: their behavior is particularly important to understand the dynamical properties of the systems when perturbed away from equilibrium, as expressed by the Kubo formula in the linear response theory~\cite{Kubo-57}. Thus, the discovery of a relation between such dynamical features and those at equilibrium, encoded in $\mathcal O(\varepsilon)$, is rather remarkable.

The second goal of the article is to show that these bounds are actually saturated in most cases of interest.
Our study will mainly be numerical and based on one-dimensional spin-$1$ quantum many-body chains.
We will consider a non-integrable tilted-field Ising model and perform an extensive study of the ETH functions for four different observables considering sizes up to $18$ sites.
We will highlight the link between the two parts of the ETH ansatz and visualize numerically how the study of autocorrelation functions, whose late-time behaviour is perfectly described by the diffusive or anomalous energy spreading hypothesis, provides the natural link between the two. 
In particular, we consider both short-range and long-range interactions which lead to, respectively, diffusive or super-diffusive spreading of the energy density. In both scenarios, our theoretical predictions are compatible with numerical computations.   
We will also present data for an observable where the bounds are not saturated, and critically discuss this issue, and describe the phenomenology for finite temperatures. 
 
\begin{figure}[t]
	\includegraphics[width=1.0\linewidth]{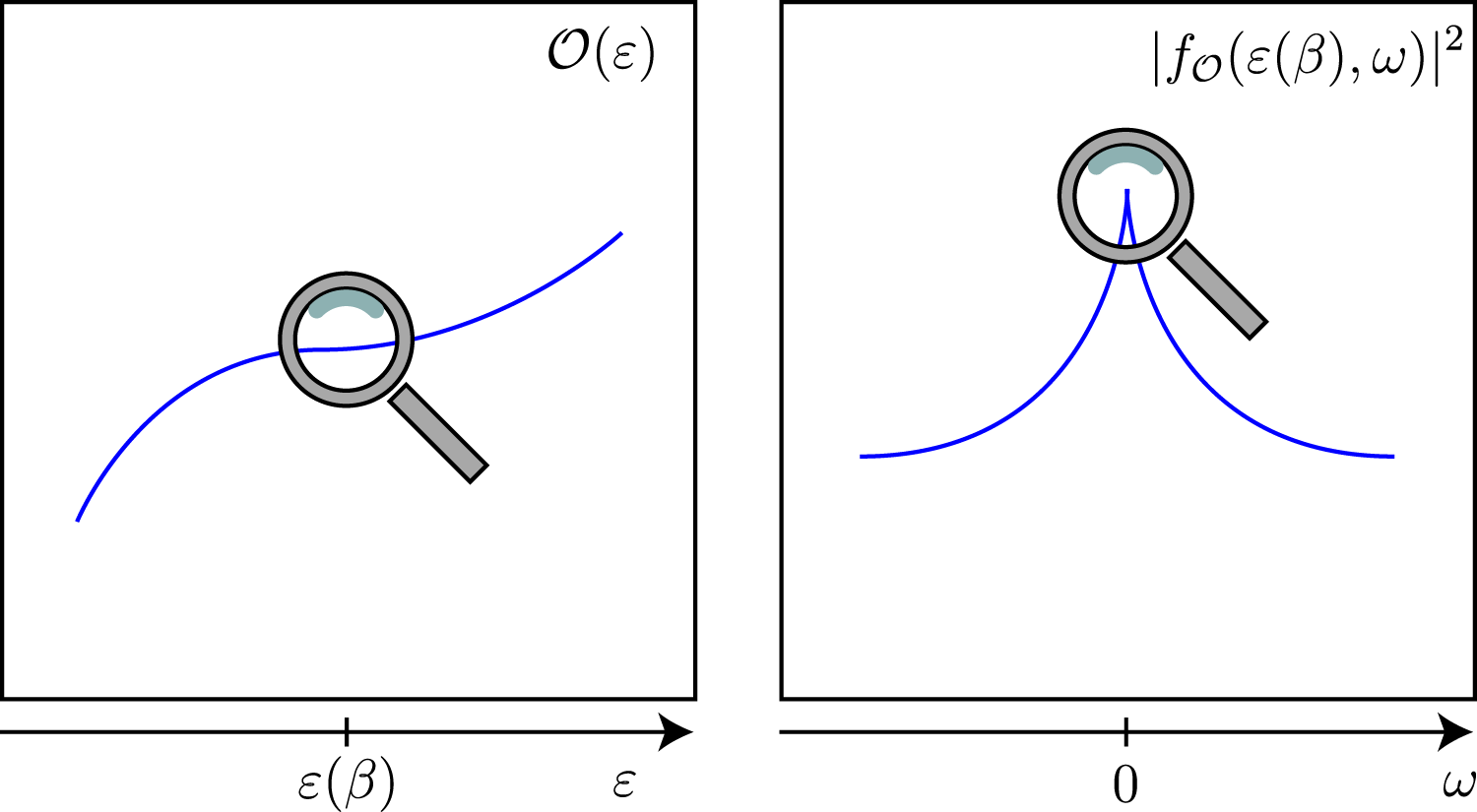}
 \caption{
The smooth profile of $\mathcal{O}(\varepsilon)$, entering the diagonal part of ETH (left), versus the behavior of $f_\mathcal{O}(\varepsilon, \omega)$ as a function of $\omega$ (right): the order of the vanishing derivative of $\mathcal{O}(\varepsilon)$ at a given energy density $\varepsilon(\beta)$ is reflected by the singularity of $|f_\mathcal{O}(\varepsilon(\beta), \omega)|^2$ in the limit $\omega \rightarrow 0$.
 }\label{fig:sketch}
\end{figure}

\subsection{Summary of results}\label{sec:summary_results}

In this subsection, we summarize in a more technical and precise way the main result of our article, namely the relation between the singularities of $f_\mathcal{O}(\varepsilon, \omega)$ in the limit $\omega \to 0$ and the profile of $\mathcal{O}(\varepsilon)$; a sketch of it is given in Fig.~\ref{fig:sketch}.
The main steps of the derivation rely on: 
(i) The identification of the notion of \textit{overlap order} of local operators, encoding their equilibrium properties; 
(ii) the expression of the autocorrelation function in terms of $f_\mathcal{O}(\varepsilon, \omega)$, and the link between the singularities at $\omega \rightarrow 0$ and the algebraic decay in time; 
(iii) a relation between the aforementioned quantities, arising from the study of the autocorrelation function through the lens of hydrodynamics.

Let us consider a one-dimensional quantum lattice model $H$ and a local observable $\mathcal{O}$, whose microcanonical expectation value at a given energy density $\varepsilon$ is given by the ETH function $\mathcal{O}(\varepsilon)$. 
In one-dimensional systems, where no phase transitions can occur at finite temperature~\cite{Araki-69},
the latter is generically a smooth function of $\varepsilon$. 

We consider an equilibrium state at inverse temperature $\beta$ with energy density $\varepsilon(\beta)$ and study the behaviour of $\mathcal{O}(\varepsilon)$ in the neighborhood of $ \varepsilon(\beta)$; under mild assumptions, it will take the form
\begin{equation}\label{eq:non_zero_der}
\mathcal{O}(\varepsilon)-\mathcal{O}(\varepsilon(\beta)) \sim (\varepsilon-\varepsilon(\beta))^m, \quad m \in \mathbb N.
\end{equation}
We call $m$ the \textit{overlap order} of $\mathcal O$, for reasons that will become clear in the main text.

Our main result is that the overlap order $m$ is intrinsically connected to the off-diagonal part of ETH. In particular, in the presence of diffusive transport, the singular behaviour of $f_\mathcal{O}(\varepsilon(\beta),\omega)$ at small $\omega$ is

\be
|f_\mathcal{O}(\varepsilon(\beta),\omega)|^2 \sim |\omega|^{m/2-1} + \text{"smooth part"},
\label{Eq:MainResult:Sec2}
\ee
as briefly summarized in Table~\ref{tab:Summary}.

We will present state-of-the-art numerics showing that for most of the observables of interest Eq.~\eqref{Eq:MainResult:Sec2} is valid: relevant exceptions are nevertheless present, and we critically discuss this issue.
A weaker result that is valid for any observable is nonetheless possible: we show that in general the divergence of $|f_{\mathcal O}|^2$ is $|\omega|^{\nu-1}$, with $\nu$ that can be extracted from the decay of autocorrelation functions,
$
    \langle \mathcal O (t) \mathcal O \rangle_c \sim {t^{-\nu}},
$ and satisfying the \textit{relaxation-overlap inequality} for diffusive transport:
\be\label{eq:ineq}
\nu \leq \frac m 2.
\ee

We show that under some mild hypotheses it is possible to exactly predict $\nu$ within a hydrodynamic framework and to identify the situations where the inequality is saturated.
The generalization of the relaxation-overlap inequality for $d$-dimensional systems having a dynamical critical exponent $z$, hence including anomalous transport, is 
\begin{equation}
\label{eq:gen_ROI}
\nu\leq \frac{dm}{z}.
\end{equation}

\begin{table}[t]
\begingroup
\renewcommand{\arraystretch}{2} 
    \begin{tabular}{ |c | c|}
    \hline
    $\mathcal{O}(\varepsilon)-\mathcal{O}(\varepsilon(\beta))$ & Singularity of $|f_\mathcal{O}(\varepsilon(\beta),\omega)|^2$\\
     \hline
     \hline
     $\sim (\varepsilon-\varepsilon(\beta))$&$\sim 1/\sqrt{\omega}$\\
    \hline
     $\sim (\varepsilon-\varepsilon(\beta))^2$&$\sim \log|\omega|$\\
     \hline
     $\sim (\varepsilon-\varepsilon(\beta))^3$&$\sim\sqrt{|\omega|}$\\
     \hline
     $\sim (\varepsilon-\varepsilon(\beta))^4$&$\sim |\omega|$\\
     \hline
     \end{tabular}
     \endgroup
 \caption{Summary of the main results under the generic assumption of diffusive spreading of energy. The behaviour $\mathcal{O}(\varepsilon)-\mathcal{O}(\varepsilon(\beta))\sim (\varepsilon-\varepsilon(\beta))^m$ of the diagonal ETH function, is reflected in the off-diagonal part of ETH. The function $|f_\mathcal{O}(\varepsilon(\beta),\omega)|^2$ develops a singularity at $\omega\rightarrow 0$ proportional to $|\omega|^{m/2-1}$, up to possible logarithmic corrections.}
 \label{tab:Summary}
\end{table}

The structure of the article is the following.
In Sec.~\ref{sec:autocorr} we derive the relaxation-overlap inequality under the assumption of one-dimensional diffusive transport ($d=1$ and $z=2$) by linking the algebraic relaxation of an autocorrelator to its saturation value at finite size. The proof is based on details that are presented in the ensuing two sections:
in Sec.~\ref{sec:fin_size}, we describe how the diagonal part of the ETH ansatz is connected to the long-time evolution of autocorrelation functions and 
in Sec.~\ref{sec:off_diag} we focus instead on its relation with the off-diagonal function $f_{\mathcal{O}}$. 
In Sec.~\ref{sec:Diffusion} we introduce the hydrodynamic approach to the spreading of energy in the system and
explain how it can be used to describe the long-time dynamics, building a connection with the ETH results in the previous sections. 
In fact, hydrodynamics has also the merit of allowing a fully general discussion of the energy spreading within quantum many-body systems, and of treating diffusion or anomalous transport on an equal footing.
The numerical study of the connection between diagonal and off-diagonal ETH functions is presented
in Sec.~\ref{sec:Numerics}, where we present extensive results on observables that reproduce the behaviours in Table~\ref{tab:Summary}. These observables
saturate the relaxation-overlap inequality, whether the Hamiltonian is short- or long-range, but we also present results for a class of operators for which the inequality is not saturated. 
Last, in Sec.~\ref{sec:conclusions} we draw our conclusions and present our outlook.

\section{Decay and saturation of the autocorrelation function}\label{sec:autocorr}    

We begin our discussion by studying the behavior of the autocorrelation function 
\begin{equation}
\la \mathcal{O}(t)\mathcal{O}\ra_{L,c} \equiv \la\mathcal{O}(t)\mathcal{O}\ra_{L}-\la \mathcal{O}\ra^2_{L}
\end{equation}
of an operator $\mathcal O$ on an equilibrium state at temperature $\beta^{-1}$ and finite size $L$. We will show that it provides an explicit link to connect the diagonal and the off-diagonal part of ETH. 

\begin{figure}[tb]
	\includegraphics[width=1.0\linewidth]{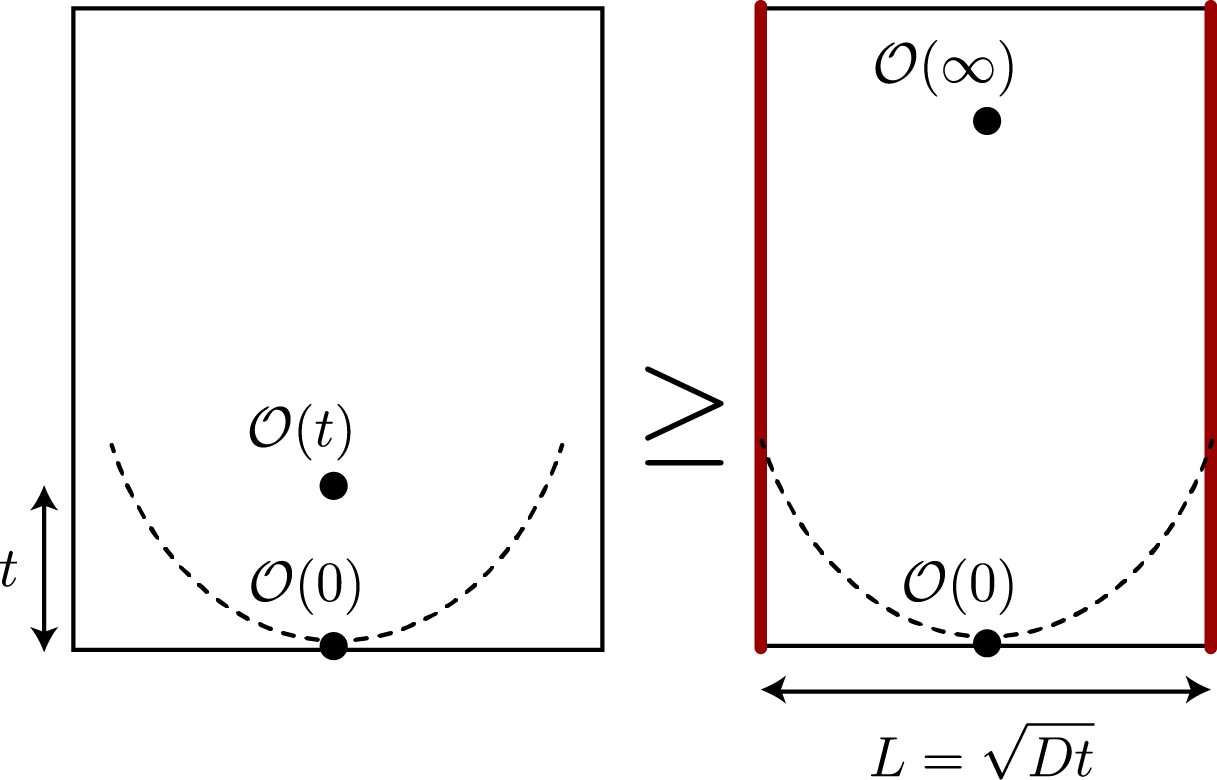}
 \caption{Comparison between the autocorrelation function of $\mathcal{O}$ in the infinite system (left) and finite-size one (right): an inequality is present and it comes from the monotonic decay in time. Due to the diffusive transport, the relevant dynamics occur inside a parabola, represented with a black dashed line. }\label{fig:corr}
\end{figure}

As anticipated in Sec.~\ref{sec:summary_results}, 
at first we will focus on a generic one-dimensional quantum lattice model, whose Hamiltonian $H$ has short-ranged interactions.
We only make the assumption that energy spreads diffusively through the system, with diffusion constant $D$.
For times not exceeding $L^2/D$,
the autocorrelation function is equal to its infinite-volume limit $L\rightarrow \infty$~\cite{bbcp-21, bsrp-22, bp-22}:
\be
\la \mathcal{O}(t)\mathcal{O}\ra_{L,c} \simeq \la \mathcal{O}(t)\mathcal{O}\ra_{
c} 
{ \qquad \text{for } \; t \lesssim \frac {L^2}{D}}
;
\ee
where $\langle \ldots \rangle_c$ denotes the autocorrelator in the infinite-volume limit.
Assuming that the decay in time of the autocorrelation function is monotonic, one can compare the two regimes above in a time-scale of order $t \lesssim \frac{L^2}{D}$ obtaining the inequality 
\be\label{eq:ineq_autocorr}
\la \mathcal{O}( t)
\mathcal{O}\ra_{c} \geq \underset{T\rightarrow \infty}{\lim}\frac{1}{T}\int^{T/2}_{-T/2} dt' \la \mathcal{O}(t')\mathcal{O}\ra_{L,c}
.
\ee
We remark that, strictly speaking, possible oscillations and revivals around the saturation value occur unavoidably, due to Poincar\'e recurrences: however, to establish the validity of Eq.~\eqref{eq:ineq_autocorr}, it is sufficient to assume that their contribution is negligible with respect to a leading monotonic profile. 
A pictorial representation of the mechanism above is given in Fig.~\ref{fig:corr}.

We now introduce two physically-motivated assumptions on the behaviours of the two members of the inequality~\eqref{eq:ineq_autocorr} that will directly lead to the relaxation-overlap inequality.
First, we assume that the autocorrelation function decays algebraically in time $t$~\footnote{We point out that, in principle, the autocorrelation function is not necessarily real, and one should refer to a monotonic decay of its absolute value. Nonetheless, it is always real for the infinite temperature state. Specifically, the complex conjugate of the autocorrelation function can be expressed as $\la \mathcal{O}(t)\mathcal{O}\ra^{*} = \la \mathcal{O}\mathcal{O}(t)\ra$ since $\mathcal{O}=\mathcal{O}^\dagger$, and, using the cyclic property of the trace, one obtains $\la \mathcal{O}\mathcal{O}(t)\ra = \la\mathcal{O}(t)\mathcal{O}\ra$ for $\beta=0$.
} 
\be\label{eq:t_decay}
\la \mathcal{O}(t)\mathcal{O}\ra_c \sim \frac{1}{t^\nu}
\ee
for a given exponent $\nu >0$, that is a common scenario in the presence of conserved charges transported through the system \cite{lmmr-14}.

Second, the right-hand side of Eq.~\eqref{eq:ineq_autocorr} probes the behaviour of the autocorrelator at finite-size and large times. 
Assuming a diffusive underlying transport the autocorrelation saturates to its equilibrium value for $t\gg L^2/D$ and in Sec. \ref{sec:fin_size} we show that
\be\label{eq:fsize_scaling}
\underset{T\rightarrow \infty}{\lim}\frac{1}{T}\int^{T/2}_{-T/2} dt \la \mathcal{O}(t)\mathcal{O}\ra_{L,c} \sim \frac{1}{L^m}, 
\ee
where $m$ is the overlap order introduced in Sec.~\ref{sec:summary_results} and that appears, for instance, in Eq.~\eqref{eq:non_zero_der}.

Finally, from \eqref{eq:ineq_autocorr}, together with \eqref{eq:fsize_scaling} and \eqref{eq:t_decay}, one gets 
$(1/L^2)^\nu \gtrsim 1/L^m$, which gives the relaxation-overlap inequality in Eq.~\eqref{eq:ineq}. 

This important result, a relation between the algebraic decays in time and space respectively, provides a way to compare diagonal and off-diagonal ETH. 
Namely, the right-hand side of the inequality \eqref{eq:ineq_autocorr}, which represents the plateau observed at finite sizes, is estimated via diagonal ETH in Sec. \ref{sec:fin_size}. 
The left-hand side, instead, is evaluated via the off-diagonal terms in Sec.~\ref{sec:off_diag}, and the algebraic decays in time manifest themselves as singularities of $f_\mathcal{O}$. This result can be extended to systems characterised by a non-diffusive energy transport leading to the generalized relaxation-overlap inequality of Eq.~(\ref{eq:gen_ROI}), that is demonstrated via hydrodynamics in Sec.~\ref{sec:Diffusion}.

\section{Finite-size saturation of the autocorrelator}\label{sec:fin_size}

In this section we characterize the right-hand side of Eq.~\eqref{eq:ineq_autocorr} and we relate it to the diagonal part of the ETH ansatz~\eqref{eq:ETH}. To wash out the off-diagonal elements of the observable $\mathcal{O}$, we first define its long-time average as
\be
\bar{\mathcal{O}} = \underset{T\rightarrow \infty}{\lim} \frac{1}{T}\int^{T/2}_{-T/2} dt \, \mathcal{O}(t).
\ee
The map $\mathcal{O}\rightarrow \bar{\mathcal{O}}$ can be thought of as a projection onto the diagonal energy-basis elements, namely $\bra{E_i}\bar{\mathcal{O}}\ket{E_j}= \bra{E_i}\mathcal{O}\ket{E_i} \delta_{ij}$; assuming ETH and non-degeneracy of the eigenspectrum, one easily gets $\bra{E_i}\bar{\mathcal{O}}\ket{E_j}\simeq \mathcal{O}(\varepsilon)\delta_{ij}$. Given $\bar{\mathcal{O}}$ we express the saturation value of the autocorrelator as
\be\label{eq:ltime_av}
\underset{T\rightarrow \infty}{\lim}\frac{1}{T}\int^{T/2}_{-T/2} dt \, \la \mathcal{O}(t)\mathcal{O}\ra_{L,c}  = \la \bar{\mathcal{O}}^2\ra_{L,c} 
\equiv 
\Delta \bar{\mathcal O}
^2_{L} \, 
\ee
where we use $\la \bar{\mathcal{O}}\mathcal{O}\ra_{L,c} = \la \bar{\mathcal{O}}^2\ra_{L,c}$; this comes from the fact that only the diagonal matrix elements enter the evaluation of these quantities, being both $\bar{\mathcal{O}}$ and the thermal density matrix diagonal in the energy basis. The variance of $\bar{\mathcal{O}}$ in the thermal state, 
noted $\Delta \bar{\mathcal{O}}^2_L$,
together with its decay with the system size, has now become the main quantity of interest.

We want to evaluate the leading order of Eq.~\eqref{eq:ltime_av} in the large system size. To do so, we regard the Hamiltonian $H$ as a Gaussian variable in the thermal state, for which we can use the well-known formula for the even moments $\langle (X- \mu)^p \rangle = \sigma^p (p-1)!!$, where $\mu$ is the mean and $\sigma^2$ the variance, and deduce the following scaling:
\begin{equation}
\label{Eq:Hamiltonian:moment:m}
    \langle \left( 
H- L \, \varepsilon(\beta)
    \right)^m \rangle_L \sim L^{\frac m2} \quad \text{for } m \text{ even}.
\end{equation}
Technically, this is legitimate because the connected cumulants of $H$ are extensive, as they can be expressed via derivatives of the thermal free energy, and therefore the central limit theorem applies. 
The fluctuations of $\bar{\mathcal{O}}$ are related to those of the energy density \cite{mv-20} and, provided that $\mathcal{O}(\varepsilon)$ is a smooth function of $\varepsilon$, we can expand at the most leading order
\be
\bar{\mathcal{O}} \simeq \mathcal{O}(\varepsilon(\beta)) + \frac{1}{m!}\frac{d^m}{d\varepsilon^m}\mathcal{O}(\varepsilon)|_{\varepsilon = \varepsilon(\beta)}\times \l \frac{H}{L} - \varepsilon(\beta) \r^m
\ee
inside the correlator, with $m$ the overlap order defined in Eq.~\eqref{eq:non_zero_der}. 
Once we do so, we observe that the variance of $\bar {\mathcal O}$ is equal to that of $\bar {\mathcal O} - \mathcal O(\varepsilon(\beta))$,
and we
express~\eqref{eq:ltime_av} at leading order as
\begin{widetext}
\begin{equation}
\Delta \bar {\mathcal O}_{L}^2
\simeq\left\la \l  \frac{1}{m!} \frac{d^m \mathcal{O}(\varepsilon)}{d\varepsilon^m} \l \frac HL-\varepsilon(\beta)\r^m\r^2\right\ra_L  
-\left\la   \frac{1}{m!} \frac{d^m \mathcal{O}(\varepsilon)}{d\varepsilon^m} \l \frac HL-\varepsilon(\beta)\r^m\right\ra^2_L\propto \frac{1}{L^m}.
\label{Eq:Fluctuations:OBar}
\end{equation}
\end{widetext}
In order to derive the last scaling, we have used Eq.~\eqref{Eq:Hamiltonian:moment:m},
that implies that both terms in Eq.~\eqref{Eq:Fluctuations:OBar} scale as $L^{-m}$; they do not cancel out exactly and the non-zero proportionality constant is computed explicitly in Appendix \ref{app:Anal_pred}.
With these few passages we have thus proven Eq.~\eqref{eq:fsize_scaling}, a result that we anticipated in Sec.~\ref{sec:autocorr}.

\subsection{Computation of the overlap order $m$ with numerical techniques}\label{SubSec:Estimating:m}

Since the exact knowledge of the function $\mathcal O(\varepsilon)$ and the reconstruction of the behaviour in Eq.~\eqref{eq:non_zero_der} for a given microscopic model can be numerically hard,
here we show a way to efficiently compute the overlap order $m$. 

We start from the observation that $\mathcal{O}(\varepsilon)$ is defined in the thermodynamic limit, where the canonical and the microcanonical ensembles are equivalent. 
Thus, assuming $\partial_\beta \varepsilon(\beta)\neq 0$, we are allowed to change variables $\varepsilon \leftrightarrow \beta$, and express the diagonal matrix elements as a function of the inverse temperature $\beta$, from now on denoted by $\mathcal{O}_\beta$. Simple algebra gives
\be\label{eq:deriv_beta}
\l-\partial_\beta\r^m\mathcal{O}_\beta = 
\l-\partial_\beta\r^m 
\frac{\text{Tr}\l e^{-\beta H}\mathcal{O}\r}{\text{Tr}\l e^{-\beta H}\r}
=\la H^m \mathcal{O}\ra_{cc},
\ee
with $\la \dots \ra_{cc}$ denoting the joint connected cumulant between $H$ and $\mathcal{O}$. 

In particular, if $m\geq 1$ is the first integer such that $\partial^m_\beta \mathcal{O}_\beta \neq 0$, 
it will also be the same for $\partial^m_\varepsilon \mathcal{O}(\varepsilon)$ via a change of variables. Moreover, since the joint moments of $H$ and $\mathcal{O}$ are functions of the connected cumulants of lower order, the first non-vanishing $\la H^m\mathcal{O}\ra_{cc}$ will also be the first non-vanishing $\la H^m\mathcal{O}\ra$, provided $\la H\ra=\la \mathcal{O}\ra=0$. This observation is what motivates the name ``overlap order'' for $m$. The remark above is beneficial for numerical applications: for example, at infinite temperature $\beta=0$, it is straightforward to evaluate
\be
\la H^m\mathcal{O}\ra|_{\beta=0} = \frac{\text{Tr}\l H^m\mathcal{O}\r}{\text{Tr}(1)},
\ee
and to find the smallest integer $m$ such that it is non-vanishing. This procedure can be generalised to a finite temperature state by computing $\langle H^m \mathcal O \rangle_\beta$ over the thermal state at non-zero inverse-temperature $\beta$.

To summarize, the overlap order $m$ entering Eq.~\eqref{eq:ineq} can be either estimated via
(i) the finite-size scaling of Eq.~\eqref{eq:fsize_scaling}, (ii)  the Taylor expansion of $\mathcal{O}(\varepsilon)$ around $\varepsilon\simeq \varepsilon(\beta)$, or
(iii) the calculation of $\langle H^m \mathcal O \rangle_{cc}$, which are from a numerical viewpoint three non-equivalent techniques.

In order to conclude this section, we discuss the specific example of $\beta = 0$: 
we assume for simplicity that $\langle H \rangle_{\beta=0}=0 $ and study
the behaviour of $\mathcal O(\varepsilon)$ around $\varepsilon \simeq 0$.
From Eq.~\eqref{eq:deriv_beta} we can easily deduce that for small $\beta$
\be
\mathcal{O}_{\beta} \simeq \frac{(-\beta)^m}{m!}\la H^m \mathcal{O}\ra_{\beta=0}+\dots
\ee
The thermal energy is similarly computed as
\be
\la H\ra_\beta \simeq (-\beta)\la H^2\ra_{\beta=0}+\dots
\ee
and therefore, we can parameterize the energy density as $\varepsilon \simeq -\beta \la H^2\ra_{\beta=0}/L$ in a neighborhood of the infinite-temperature state. 
By inverting the relation between $\varepsilon$ and $\beta$ that we just obtained,
we express the ETH function as
\be
\mathcal{O}(\varepsilon)\simeq 
\frac{\varepsilon^m}{m!} \frac{\la H^m \mathcal{O}\ra_{\beta=0}}{(\la H^2\ra_{\beta=0} /L)^m} +\dots
\label{eq:fit_formula} 
\ee
The expression that we just derived can be used to interpret the numerical data obtained for estimating the ETH function $\mathcal O(\varepsilon)$ as it gives a quantitative prediction for its lowest-order Taylor expansion around $\varepsilon = 0$.

\section{Singularities and algebraic relaxation}\label{sec:off_diag}

In this section we characterize the left-hand side of Eq.~\eqref{eq:ineq_autocorr} and we relate it to the off-diagonal part of the ETH ansatz~\eqref{eq:ETH}.
The connected autocorrelation function of a local observable $\mathcal{O}$ is expressed in the thermodynamic limit ($L\rightarrow \infty$) as~\cite{dkpr-16,ms-19,sjhv-21}
\begin{align}
\label{eq:2pt_ETH1}
\la \mathcal{O}(t)\mathcal{O} \ra_{c} = \int d\omega e^{-\beta\omega/2+i\omega t} |f_\mathcal{O}\l \varepsilon(\beta),\omega\r|^2. 
\end{align}
The function $f_{\mathcal O}(\varepsilon, \omega)$ is crucial in our discussion and several of its properties have already been characterised.
For instance, at large $\omega$, the function $f_\mathcal{O}$ is known to decay rapidly~\cite{ms-19} in order to let the integral converge notwithstanding the exponential term $e^{- \beta \omega /2}$. 
Instead, important finite-size effects are present at small frequencies $\omega \lesssim 1/L^2$, 
intuitively related to times $t \sim L^2$ that characterize the spreading on the entire system of a diffusive process and that are typically described using Random Matrix Theory~\cite{dkpr-16,np-18,bgr-20}.
Here, since we are interested in the $L\to \infty$ limit, these small-frequency effects will be completely disregarded.
Other bounds on $f_{\mathcal O}(\varepsilon, \omega)$ have been presented in Ref.~\cite{nc-22}.

We now discuss the relation between the late-time autocorrelator at infinite size in Eq.~\eqref{eq:2pt_ETH1} and the singularities of $f_\mathcal{O} (\varepsilon, \omega)$ at small frequencies. To do so, 
it is first worth recalling a few results about Fourier transform, found in textbooks~\cite{bc-49}. 
Let us consider  a function $F(t)$ and  its Fourier transform $\hat{F}(\omega)$
\be
F(t) = \int^{+\infty}_{-\infty} d\omega \, e^{i\omega t}\hat{F}(\omega).
\ee
The Riemann-Lebesgue theorem 
ensures that, if $\hat{F} \in L^1(\mathbb{R})$, 
namely the integral of its absolute value is finite, 
then
\be
\underset{t\rightarrow \infty}{\lim} F(t) =0.
\ee
If $\hat{F}(\omega)$ is smooth, namely $\hat{F} \in C^{\infty}(\mathbb{R})$, using the Riemann-Lebesgue theorem and integration by parts,  one can further prove that $F(t)$ goes to zero faster than any power law as $t\rightarrow \infty$. 
More precisely, a bound for any integer $n$ exists using a properly chosen constant $C_n$:
\begin{equation}
|F(t)|\leq \frac{C_n}{|t|^n},
\qquad \forall n \in \mathbb N,
\quad \forall t \in \mathbb R.
\end{equation}

From these general results, we learn that the convergence of the integral of $|f_\mathcal{O}\l \varepsilon,\omega\r|^2$ as a function of $\omega$ ensures the decay to zero of the autocorrelation function~\eqref{eq:2pt_ETH1} at large times. 
On the other hand, the algebraic decay in~\eqref{eq:t_decay} signals that $|f_\mathcal{O} (\varepsilon, \omega)|^2 $ cannot be smooth as a function of $\omega$, and discontinuities in the derivatives must arise.

For the sake of clarity, we focus on $\beta=0$, so that $|f_\mathcal{O}(\varepsilon,\omega)|^2$ 
is the Fourier transform of the autocorrelation function of $\mathcal{O}$, and in this discussion we drop the explicit dependence on $\varepsilon$. 
The study of a specific $\beta$ does not imply a loss of generality, since the contribution coming from $\beta$ in \eqref{eq:2pt_ETH1} is the smooth function $e^{-\beta \omega/2}$ and its presence does not affect the singularities in $\omega$. We also point out that, from the definition \eqref{eq:ETH}, the function $f_{\mathcal{O}}(\omega)$ satisfies  $|f_\mathcal{O}(\omega)|^2=|f_\mathcal{O}(-\omega)|^2$, which forbids a certain class of discontinuities. 

\begin{table}[t]
\begingroup
\renewcommand{\arraystretch}{2} 
    \begin{tabular}{ |c | c|}
    \hline
    $\langle \mathcal O(t) \mathcal O \rangle_c $ & Singularity of $|f_\mathcal{O}(\varepsilon(\beta), \omega)|^2$\\
     \hline
     \hline
     $\sim t^{-\frac 12}$&$\sim 1/\sqrt{\omega}$\\
    \hline
     $\sim t^{-1}$&$\sim \log|\omega|$\\
     \hline
     $\sim t^{-\frac 32}$&$\sim\sqrt{|\omega|}$\\
     \hline
$\sim t^{- 2}$&$\sim |\omega|$\\
     \hline
     \end{tabular}
     \endgroup
 \caption{Relations between the algebraic decay of the autocorrelation function of the thermal state at inverse temperature $\beta$ and the singularities of $f_{\mathcal O}(\varepsilon(\beta), \omega)$ at small $\omega$.}
 \label{tab:nu}
\end{table}

From scaling arguments, we infer a singular behaviour of $f_\mathcal{O} (\omega) $ associated with the algebraic decay in time of the autocorrelation function $\sim t^{- \nu}$ in Eq.~\eqref{eq:t_decay} as follows:
\be
|f_\mathcal{O}(\omega)|^2 \sim |\omega|^{\nu-1} + \text{"smooth part"}.
\ee
Possible logarithmic corrections might arise in principle in the singular part, without changing the leading behaviour; 
in particular, if $\nu$ is an odd integer, they should surely arise because  $|\omega|^{\nu-1} = \omega^{\nu-1}$ is a smooth function:
\begin{equation}
|f_\mathcal{O}(\omega)|^2 \sim \omega^{\nu-1}\log|\omega| + \text{"smooth part"},
\quad \text{for } \nu \text{ odd.}
\end{equation}
In Table~\ref{tab:nu}
we present some examples that will be useful for the interpretation of the numerical simulations of Sec.~\ref{sec:Numerics}.

It is worth noting that additional singularities at $\omega\neq 0$ might be present in principle as well. However, these are associated with oscillations whose amplitude decays algebraically.
For example, if $|f_\mathcal{O}(\omega)|^2 \sim |\omega-\omega_0|^{\nu-1}+\text{"smooth part"}$ for $\omega \simeq \omega_0 \neq 0$, a contribution $\cos(\omega_0 t)/t^{\nu}$ is present in the autocorrelation function. 
We do not focus on these effects since they can be regarded as transient oscillations on top of the leading and monotonically decaying profile of the autocorrelation function.

\subsection{Time derivatives of local operators}\label{sec:t_der}

At this stage, we have derived all relations that are necessary to deduce the relaxation-overlap inequality in Eq.~\eqref{eq:ineq} using the arguments explained in Sec.~\ref{sec:autocorr}. We stress that this inequality is not necessarily saturated and now we provide a relevant counterexample: 
given a generic local operator $\mathcal{O}$, we consider its time derivative as
\be
\label{Eq:Oprime}
\mathcal{O}' \equiv \frac{d}{dt}\mathcal{O}(t) = i[H,\mathcal{O}].
\ee
We begin by discussing the inequality in Eq.~\eqref{eq:ineq_autocorr} and we show that for $\mathcal O'$ it is not saturated.
Assuming ETH for $\mathcal{O}$, we easily compute from Eq.~\eqref{eq:ETH} that ETH holds for $\mathcal{O}'$ as well:
\be
\bra{E_i}\mathcal{O}'\ket{E_j} \simeq  i\, \omega \, f_\mathcal{O}(\varepsilon,\omega)\exp\l -S(\varepsilon)/2\r R_{ij}.
\ee  
In particular, the diagonal part is vanishing, namely $\mathcal{O}'(\varepsilon)=0$.

Let us now show that for this operator the bound cannot be saturated. To do so, we observe that
\be\label{eq:sec_derivative}
\la \mathcal{O}'(t)\mathcal{O}'\ra_c = \frac{d^2}{dt^2}\la\mathcal{O}(t)\mathcal{O}\ra_c,
\ee
and assuming a decay $\la \mathcal{O}(t)\mathcal{O}\ra_c \sim t^{-\nu}$ for $\mathcal{O}$, algebraic relaxation has to be present for $\mathcal{O}'$, and it holds
\be
\la \mathcal{O}'(t)\mathcal{O}'\ra_c \sim \frac{1}{t^{\nu+2}}.
\ee

On the other hand, we can look at these results from the different perspective of the relaxation-overlap inequality~\eqref{eq:ineq}.
Concerning the size-dependence of the autocorrelation function, it is identically equal to $0$ at any finite size, since from Eq.~\eqref{eq:sec_derivative} one has
\be
\underset{T\rightarrow \infty}{\lim}\frac{1}{T}\int^{T/2}_{-T/2} dt \, \la \mathcal{O}'(t)\mathcal{O}'\ra_{L,c} =0.
\ee
For instance,~\eqref{eq:sec_derivative} is a total derivative, and it only gives boundary terms when integrated over $t$; since those are finite, their ratio with $T$ is vanishing for $T \rightarrow \infty$.
Thus, the quantity above is smaller than every power law $1/L^{m }$, and therefore we identify the overlap order of $\mathcal{O}'$ as $m'=\infty$.
We can recover the same result by checking that 
the connected cumulants of $\mathcal O'$ with the powers of the Hamiltonian, $\la H^{m} \mathcal{O}'\ra_c$, are identically zero for any $m \geq 1$.

Let us now inspect the off-diagonal part of the ETH for $\mathcal O'$.
We can relate the off-diagonal elements of $\mathcal{O}'$ to those of $\mathcal{O}$ as follows
\be\label{eq:fO1}
|f_{\mathcal{O}'}(\omega)|^2 = \omega^2|f_{\mathcal{O}}(\omega)|^2.
\ee
Thus, singularities of $|f_{\mathcal{O}}(\omega)|^2$ manifests as singularities in $|f_{\mathcal{O}'}(\omega)|^2$ as well. We remark that Eq.~\eqref{eq:fO1} predicts $|f_{\mathcal{O}'}(\omega\rightarrow 0)|^2 =0$, even in the worst case scenario where $f_{\mathcal{O}}$ diverges as $|f_{\mathcal{O}}(\omega)|^2\sim 1/\sqrt{\omega}$.
This means that the singularities in $\omega$ imply a power-law decay as $t^{- \nu'}$ with $\nu' = \nu+2$. The inequality~\eqref{eq:ineq}, that in this case reads $\nu' \leq m'/2$, is thus satisfied but not saturated.

\section{Hydrodynamics and the relaxation-overlap inequality}\label{sec:Diffusion}

In this section, we first provide a perspective based on hydrodynamics on the relaxation-overlap inequality in Eq.~\eqref{eq:ineq}, which is obtained under the assumption that energy spreads diffusively in the system.
We will see that it is possible to derive it using only the standard hydrodynamic framework that is usually employed for closed quantum many-body systems. Strengthened by this result,
we will address a second goal, namely that of extending the hydrodynamic theory to include the possibility of anomalous diffusion, and derive the general relaxation-overlap inequality in Eq.~\eqref{eq:gen_ROI}.

\subsection{Hydrodynamic diffusion theory}\label{sec:diffusive:hydro:theory}

We assume that the Hamiltonian (energy) density $h(x)$ is the only local conserved charge of the model, and further assume that it spreads diffusively~\footnote{
While the local Hamiltonian $h(x)$ is defined up to a gauge redundancy $h(x)\rightarrow h(x)+\partial_x \Lambda(x)$, with $\Lambda(x)$ any local operator, the hydrodynamic equation \eqref{eq:diffusion} is only consistent in specific gauges (see Ref.~\cite{dbd-19} for further details).
} according to the well-known equation:
\be\label{eq:diffusion}
\partial_t h(x,t) \simeq D \, \partial^2_x h(x,t),
\ee
with $D$ the diffusion constant.

From Eq.~\eqref{eq:diffusion}, one can systematically study the time evolution of the higher-order correlation function of the Hamiltonian density. Moreover, assuming the hydrodynamic principle, 
one can compute the autocorrelation function of any local operator $\mathcal{O}(x)$
by projecting it onto $h(x)$ and its powers to get the most leading contributions to its relaxation at large time. 
Naively, if $\mathcal{O}(x)$ is an observable with support around a given position $x$, the formal expansion reads
\be\label{eq:naive_proj}
\mathcal{O}(x) \sim h(x)+\partial_x h(x)+\dots +\normOrd{h(x)^2}+\normOrd{h(x)\partial_x h(x)}+\dots,
\ee
with some proper numerical prefactors that we have omitted. The symbol $\normOrd{h(x)^m}$ refers to a \textit{normal ordering} procedure, where the projections onto smaller powers of $h(x)$ are subtracted from $h(x)^m$; we give more details in Appendix~\ref{app:proj}. 
From the expansion~\eqref{eq:naive_proj} and the hydrodynamic equation~\eqref{eq:diffusion}, an algebraic decay in time of the autocorrelation function of $\mathcal{O}(x)$ can be derived, as we show below.

Let us begin by considering the algebraic decay of the autocorrelation function of the operator $\normOrd{h(x)^n}$. To do so, it is convenient to investigate the correlator
\be\label{eq:Correlator}
C(\mathbf{x}, t) = C(x_1,\dots,x_n;t)
\equiv \la \normOrd{h(x_1,t)\dots h(x_n,t)}\mathcal{O}'\ra
\ee
with $\mathcal{O}'$ any local observable, for generic $\mathbf x$, to eventually recover the case of interest by setting all the coordinates $x_j \rightarrow x$. 
Before entering into the details of the derivation, we want to stress that there are two main reasons to use the normal-ordered $n$-th power of the Hamiltonian density. First, possible issues arising from insertions of $h(x,t)$ at close-by points, which might locally spoil the validity of the diffusion equation, are expected to be washed out. Second, for $|x_j|\rightarrow \infty$ the function \eqref{eq:Correlator} goes to zero since disconnected terms are explicitly subtracted and the thermal state $\la \dots\ra$ satisfies clustering properties (that is, the connected correlation functions decay exponentially in the distance \cite{Araki-69}); therefore, its spatial Fourier transform is well-behaved, and it is a smooth function of the momentum. 
 
Assuming diffusion from Eq.~\eqref{eq:diffusion}, the following differential equation holds
\be\label{eq:diff_eq}
\partial_t C
(\mathbf{x}, t)
= D(\partial^2_{x_1}+\dots+\partial^2_{x_n})C(\mathbf{x}, t)
,
\ee
which determines unambiguously $C(\mathbf{x}, t)$ once its value at $t=0$ is fixed. 
In particular, we are allowed to express the solution of Eq.~\eqref{eq:diff_eq} in Fourier space as
\be\label{eq:Solution_diff}
C(\mathbf{x},t) = \int \frac{d^n\mathbf{k}}{(2\pi)^n}\tilde{C}(\mathbf{k},t=0)e^{-D\mathbf{k}^2t}e^{i\mathbf{k\cdot x}},
\ee
with
\be
\tilde{C}(\mathbf{k},t=0) \equiv \int d^n \mathbf{x} e^{-i\mathbf{k\cdot x}}C(\mathbf{x},t=0).
\ee
The late-time behaviour of Eq.~\eqref{eq:Solution_diff} at fixed $\mathbf{x}$ can be computed by saddle-point analysis around $\mathbf{k}=0$ and it gives
\be\label{eq:C_decay}
C(\mathbf{x},t)\sim \frac{\tilde{C}(\mathbf{k}=0,t=0)}{t^{n/2}},
\ee
up to an irrelevant prefactor.

The result above is compatible with the relaxation-overlap inequality in Eq.~\eqref{eq:ineq}. 
For instance, if $\mathcal{O}(x)\simeq \normOrd{h(x)^n}+\dots$, the projections of $\mathcal{O}$ onto the powers of $H$ with degree smaller than $n$ vanish. 
On the other hand, the projection of $\mathcal O(x)$ onto $H^n$ is non-vanishing. Thus, $n$ is the overlap order of the operator as discussed in Sec.~\ref{sec:fin_size}. 
From what we just presented, the most leading contribution to the time-decay of the autocorrelator is $\la \mathcal{O}(t)\mathcal{O}\ra\sim t^{-n/2}$:
we have thus found that $\nu = m/2$, which saturates the relaxation-overlap inequality~\eqref{eq:ineq}.

Exceptions to the scaling in Eq.~\eqref{eq:C_decay} are found whenever $\tilde{C}(\mathbf{k}=0,t=0)=0$. In that case, we can still observe a power-law decay with an exponent different from $n/2$. For example, if $\tilde{C}(\mathbf{k},t=0)\sim |\mathbf{k}|^\mu$ for small $\mu$, a scaling argument gives
\be\label{eq:C_nu_mu}
C(\mathbf{x},t) \sim \frac{1}{t^{(n+\mu)/2}}.
\ee
An explicit example is provided by the correlation function of $\la \partial_t h(x,t)\partial_t h(0,0)\ra$, the operator that we have analysed in Sec.~\ref{sec:t_der}, where we have shown with different arguments that it does not saturate the inequality in Eq.~\eqref{eq:ineq}.
Indeed, from a hydrodynamic perspective, it satisfies a diffusion equation, but according to Eq.~\eqref{eq:sec_derivative} its late-time behavior is 
\be\label{eq:dth_decay}
\la \partial_t h(x,t)\partial_t h(0,0)\ra = \partial_t^2 \la h(x,t)h(0,0)\ra \sim \frac{1}{t^{5/2}}.
\ee
The scaling above can be equivalently obtained from the representation of the correlation function in Fourier space in the right-hand side of Eq.~\eqref{eq:Solution_diff}. For instance, if we take the second time-derivative of the autocorrelation function of $h(x,t)$, an additional factor $|k|^4$ appears inside the integral of~\eqref{eq:Solution_diff}, and $\mu=4$ for the operator $\partial_t h$: therefore, since $n=1$ here, we have from Eq.~\eqref{eq:C_nu_mu} the decay $\sim t^{-(n+\mu)/2} = t^{-5/2}$. Similarly, one could show that
\be
\la \partial_x h(x,t)\partial_x h(0,0)\ra = \partial^2_x \la h(x,t) h(0,0)\ra \sim \frac{1}{t^{3/2}},
\ee
and the Fourier transform of the autocorrelation function of $\partial_x h$ goes as $k^2$ (corresponding to $\mu=2$) for small $k$.

In general, operators that are space- or time-derivatives of local operators have exactly vanishing projections onto the powers of the global Hamiltonian $H$, and, consequently, vanishing diagonal elements; we have discussed this in detail in Sec.~\ref{sec:t_der}. For these operators, the bound in Eq.~\eqref{eq:ineq} becomes trivial, as $m $ tends to infinity and the inequality is never saturated. Similarly, these operators do not satisfy the late-time scaling in Eq.~\eqref{eq:C_decay}. However, hydrodynamics allows going beyond that and understanding the algebraic relaxation of those operators as well.

Finally, it is worth analysing the finite-size effects predicted by diffusion, with an emphasis on the infinite-time saturation value of the correlators. To do so, we assume the validity of Eq.~\eqref{eq:diff_eq} in a system of size $L$. 
This amounts to quantizing the momenta, with a quantization condition that depends on the boundary conditions (for example periodic or open), and the formal solution to $C(\mathbf{x},t)$ can be obtained by replacing
\be
\int \frac{d^n\mathbf{k}}{(2\pi)^n} \rightarrow  \frac{1}{L^n}\sum_{\mathbf{k}}
\ee
in Eq. \eqref{eq:Solution_diff}. At infinite time, one gets
\be
C(\mathbf{x},t=+\infty) = \frac{\tilde{C}(\mathbf{k}=0,t=0)}{L^n},
\ee
which is the same scaling found in Sec.~\ref{sec:fin_size}. Therefore, we conclude that hydrodynamics predicts that the exponent entering the finite-size scaling of the asymptotic value of the autocorrelator is precisely the overlap order. This result complements the same finding obtained in Sec.~\ref{sec:autocorr}, which makes use of the ETH.

It is worth noting that this finite-size effect is extremely sensitive to the boundary conditions, and whether $\mathbf{k}=0$ belongs to the set of quantized momenta. 
Also, if $\tilde{C}(\mathbf{k}=0,t=0)=0$, then $C(\mathbf{x},t=+\infty)$ vanishes identically, and for example it does not decay with a power of $1/L$. 
These last observations point out explicitly that one should be extremely careful about the exchange $L \leftrightarrow Dt^2$, which is the reason why the inequality~\eqref{eq:ineq_autocorr} is not necessarily saturated.

Another remark is that, given an observable $\mathcal{O}$, its formal expansion~\eqref{eq:naive_proj}, as well as the notion of overlap order, depends on the temperature. Thus, its autocorrelator could decay as $\sim t^{-m/2}$ at a specific temperature, while it behaves as $\sim t^{-1/2}$ away from it (as the overlap with the energy density becomes non-vanishing). This mechanism is explicitly realized in the numerical examples of Sec.~\ref{sec:Numerics}, where a comparison between $\beta =0$ and $\beta\neq 0$ is presented.

\subsection{Higher dimensions and anomalous transport}\label{sec:generaliz}

The algebraic scalings in time discussed so far refer to one-dimensional systems with an underlying diffusive transport. We ask whether similar results can obtained for a $d$-dimensional system and in the presence of anomalous transport.

We first point out that, assuming the extensivity of the Hamiltonian and diagonal ETH, one can generalize the discussion in Sec.~\ref{sec:fin_size}; the saturation value of the autocorrelator of $\mathcal{O}$ now reads
\be\label{eq-Ldm}
\underset{T\rightarrow \infty}{\lim}\frac{1}{T}\int^{T/2}_{-T/2} dt \la \mathcal{O}(t)\mathcal{O}\ra_{L,c} \sim \frac{1}{L^{dm}},
\ee
with $L^d$ the system's volume. Furthermore, if the critical exponent associated with the transport is $z$ one expects the saturation to occur at a time scale $t\sim L^z$. From Eq. \eqref{eq-Ldm} and the inequality \eqref{eq:ineq_autocorr}, based on the monotonic behavior of the autocorrelation function, one shows that the exponent $\nu$ entering the algebraic decay $
\la \mathcal{O}(t)\mathcal{O}\ra \sim {t^{-\nu}},
$ must satisfy Eq.~(\ref{eq:gen_ROI}): the general relaxation-overlap inequality is thus proven within this conventional hydrodynamics framework. 
We note that Eq.~\eqref{eq:ineq} is recovered for $d=1$ and $z=2$.

Paradigmatic examples where the non-diffusive scaling $z \neq 2$ is found are long-range systems~\cite{slk-20,Knap-22,ddmprt-23}. 
The anomalous diffusion of the conserved charges, for instance of the Hamiltonian density $h(x)$, is driven by the fractional Laplacian \cite{Catalano-23} as 
\be
\partial_t h(x,t) = D \nabla^{z} h(x,t).
\ee
In that case, the calculations of Sec.~\ref{sec:diffusive:hydro:theory} for the diffusive case can be generalized and we find:
\be
\la \normOrd{h(x,t)^m} \normOrd{h(0,t)^m}\ra \sim \begin{cases}
t^{-md/z}, & \text{for } t\lesssim L^{z};\\ L^{-md}, & \text{for } t\rightarrow \infty.
\end{cases}
\ee

Finally, we mention that ballistic transport ($z=1$) in short-range systems is usually associated with integrability and an infinite number of conserved charges. Thus, a proper treatment of the transport in that regime requires a more general framework as generalized hydrodynamics (GHD)~\cite{cdy-16, bcdf-16}, we leave this for future work.

\section{Numerical analysis}\label{sec:Numerics}

\begin{figure}[t]
	\includegraphics[width=1.0\linewidth]{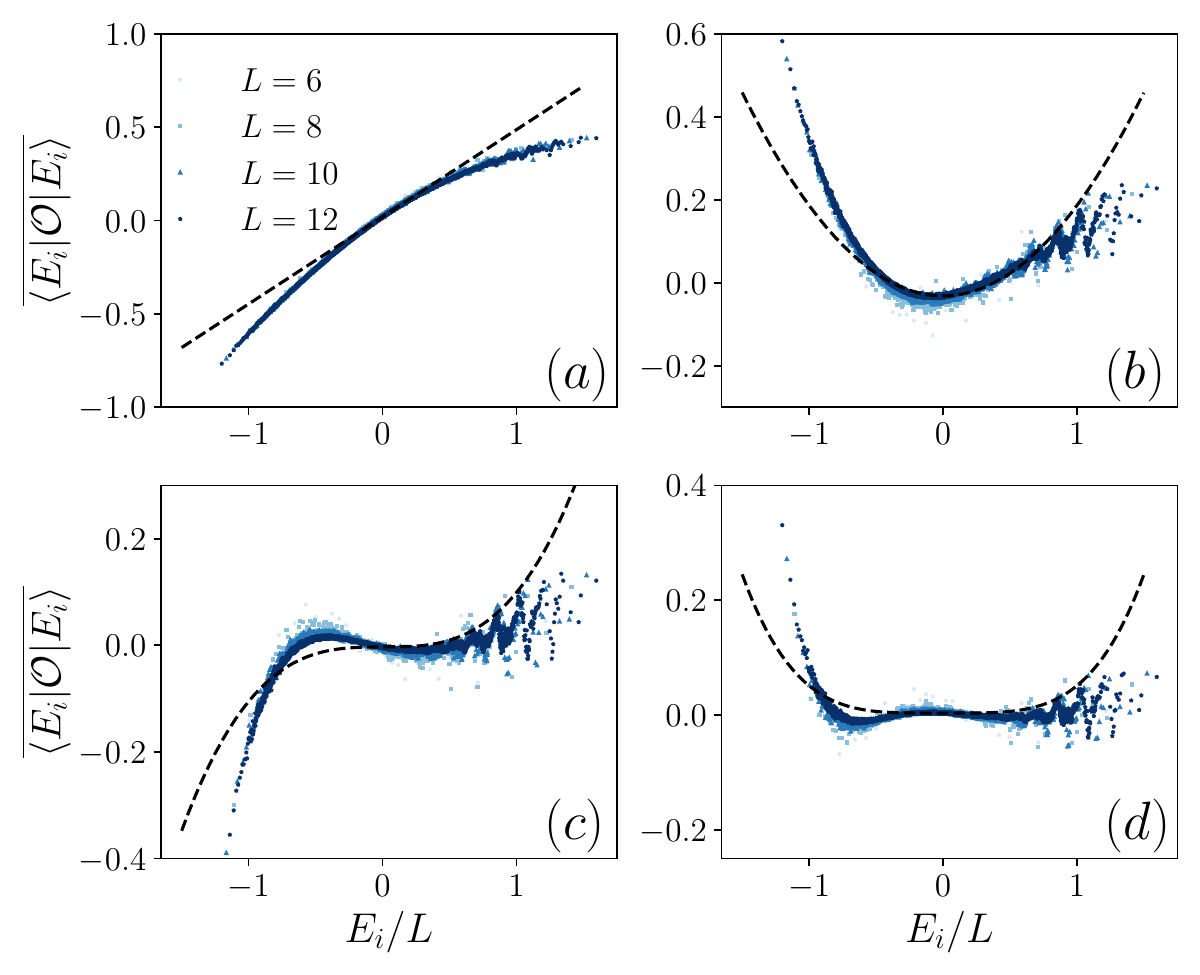}
 \caption{Diagonal matrix elements as a function of the energy density in the tilted field spin-1 Ising model for the operators (a) $S_{1}^{x}$; (b) $S_{1}^{x}S_{2}^{x}$; (c) $S_{1}^{x}S_{2}^{x}S_{3}^{x}$; (d) $S_{1}^{x}S_{2}^{x}S_{3}^{x}S_{4}^{x}$, for system size $L=6,8,10,12$, where an additional moving average is taken over $27$ eigenstates to suppress fluctuations. The dashed lines show the scaling  $\mathcal{O}(\varepsilon)-\mathcal{O}(\varepsilon=0)\sim \varepsilon^m$ where $\varepsilon = \frac{E_i}{L}$, and the prefactor is calculated using Eq.~\eqref{eq:fit_formula}. }\label{Fig-DETH}
\end{figure}

\begin{figure}[tb]
	\includegraphics[width=1.0\linewidth]{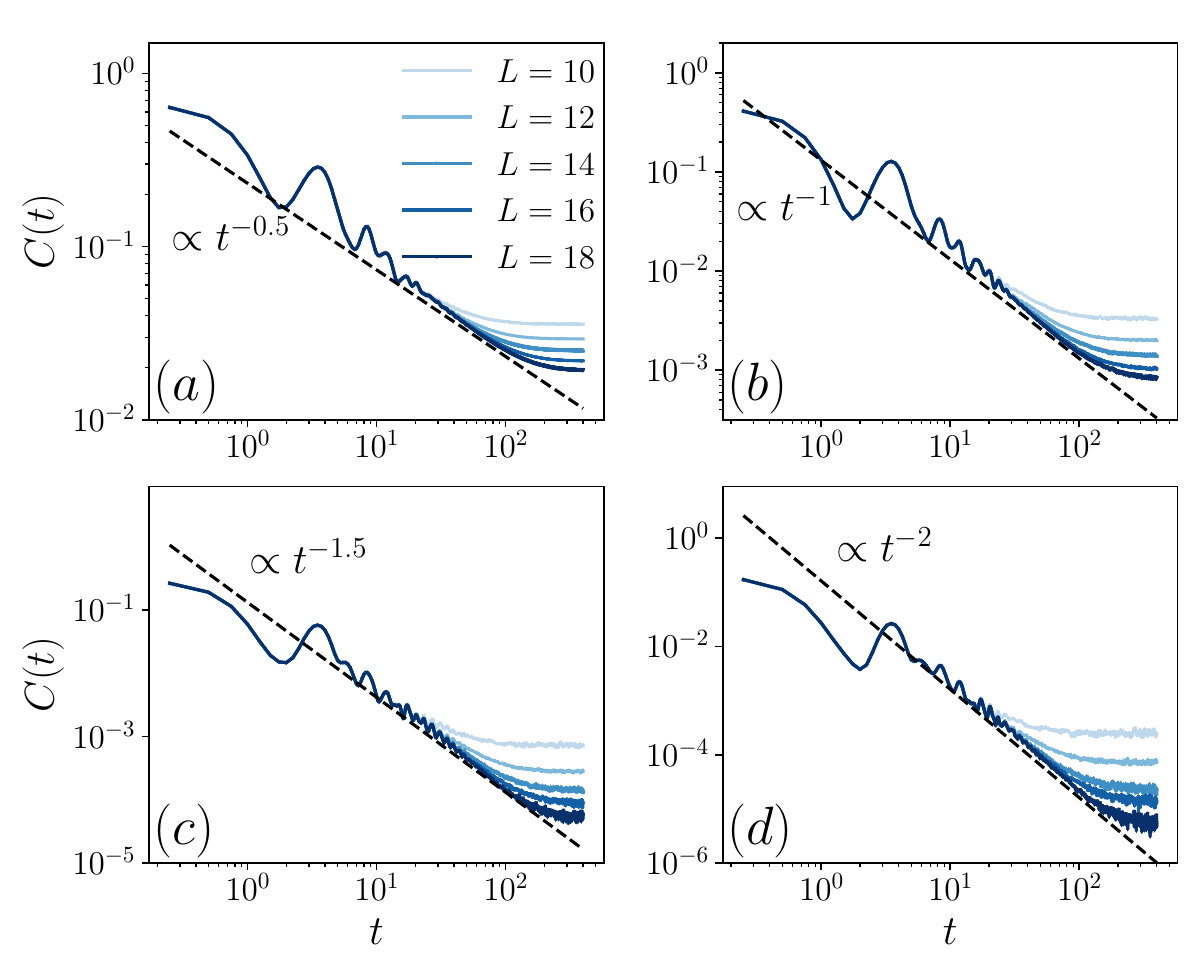}
 \caption{Autocorrelation function $C(t)$ at infinite temperature in the tilted field spin-1 Ising model for the operators (a) $S_{1}^{x}$; (b) $S_{1}^{x}S_{2}^{x}$; (c) $S_{1}^{x}S_{2}^{x}S_{3}^{x}$; (d) $S_{1}^{x}S_{2}^{x}S_{3}^{x}S_{4}^{x}$, for system sizes $L=10, 12,14,16,18$. The dashed lines indicate the expected scaling $\propto 1/t^{m/2}$. Results for $L=10$ are calculated by exact diagonalization, and for $L > 10$ results are calculated using DQT.}\label{Fig-AAT}
\end{figure}

\begin{figure}[tb]
\includegraphics[width=1.0\linewidth]{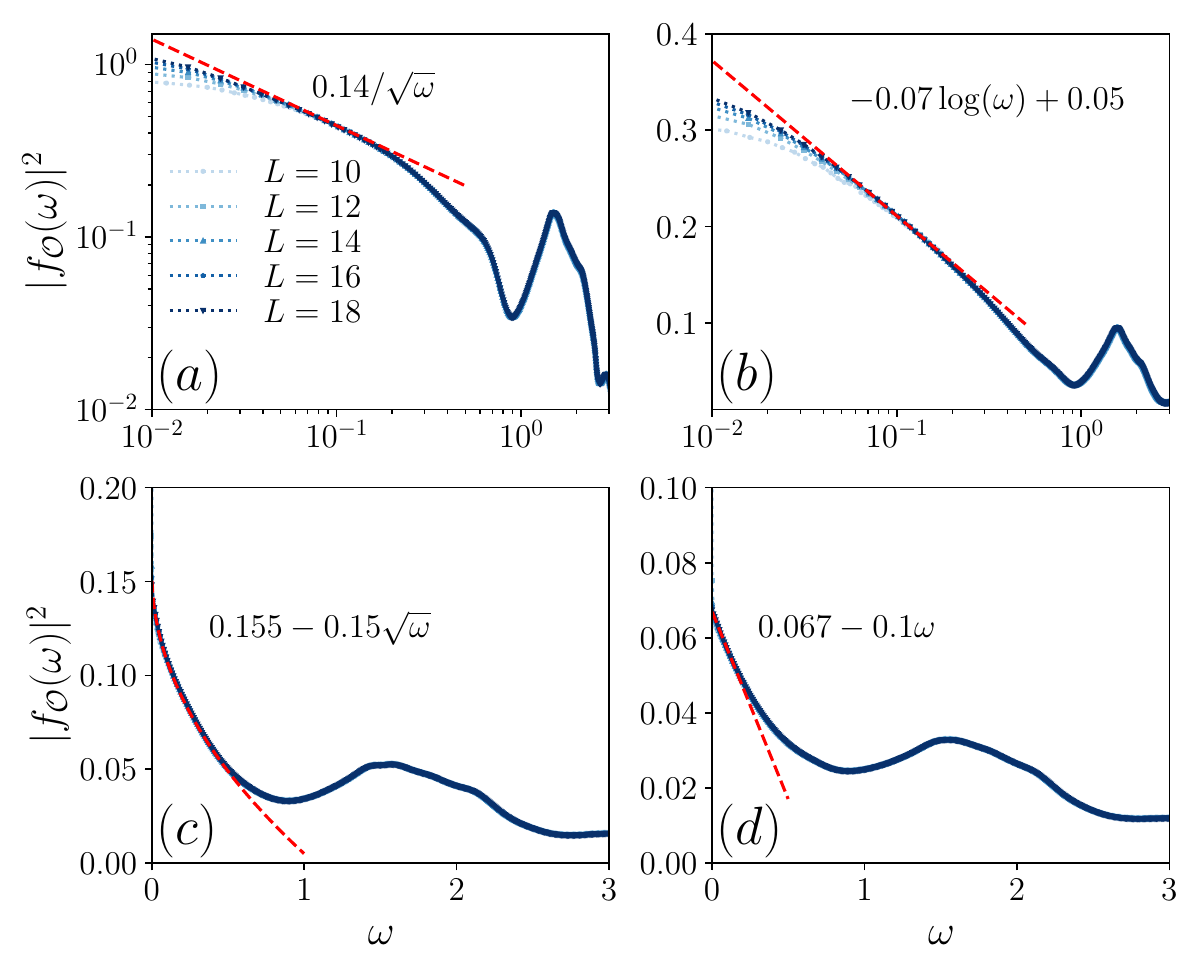}
 \caption{$|f_{\cal O}(\omega)|^2$ versus $\omega$, in the tilted field spin-1 Ising model for operators (a) $S_{1}^{x}$; (b) $S_{1}^{x}S_{2}^{x}$; (c) $S_{1}^{x}S_{2}^{x}S_{3}^{x}$; (d) $S_{1}^{x}S_{2}^{x}S_{3}^{x}S_{4}^{x}$, for system size $L=10,12,14,16,18$, for $\beta = 0.0$. Results for $L = 10$ are calculated by ED, while results for $L>10$ are calculated by the Fourier transform of the corresponding autocorrelation functions shown in Fig.~\ref{Fig-AAT} computed by DQT. The dashed line gives the singular behaviour in panels (a) to (d) ($\propto 1/\sqrt{\omega}, \log(\omega), \sqrt{\omega}, \omega$), respectively.}\label{Fig-FW}
\end{figure}

We now present a series of numerical simulations for one-dimensional spin models which can be interpreted using the theoretical tools developed so far. 
We will employ these simulations to validate the relation between diagonal and off-diagonal ETH functions that we have summarized in Table~\ref{tab:Summary},
which is our main result.
We will also discuss the numerical data from the viewpoint of the
relaxation-overlap inequality in Eq.~\eqref{eq:ineq},  
and show that in many cases of interest it is actually saturated. 
We have already identified in Sec.~\ref{sec:Diffusion} the situations where the inequality is saturated, 
yet the discussion relied on the hypothesis of performing a hydrodynamic study;
discussing Eq.~\eqref{eq:ineq} from a numerical viewpoint constitutes thus an unavoidable passage.

As a first model, we consider a spin-$1$ Ising chain with tilted field
\begin{equation}\label{eq:H_Ising}
H=\sum_{l=1}^{L}(JS_{l}^{z}S_{l+1}^{z}+h_{x}S_{l}^{x}+h_{z}S_{l}^{z});
\end{equation}
we set periodic boundary conditions and $h_x = 1.1$, $h_z = 0.9$, and $J = 0.707$. 
Here $S^\mu_l$, with $\mu \in \{x,y,z\}$, are spin-$1$ operators at site $l$ which fulfill the usual on-site commutation relations $[S^x_l,S^y_l] = i S^z_l$. 
For the parameters chosen above, the model is non-integrable and exhibits diffusive transport.
We focus on the infinite temperature state ($\beta=0$) and on its autocorrelation functions, analyzing the finite temperature case in Sec.~\ref{sec:finiteT};
any thermal average simply requires
to consider $\la \dots \ra \equiv \text{Tr}(\dots)/\text{Tr}(1)$. 
We pick the following set of observables, $S_l^x$, $S^x_{l}S^x_{l+1}$, $S_{l}^{x}S_{l+1}^{x}S_{l+2}^{x}$ and $S_{l}^{x}S_{l+1}^{x}S_{l+2}^{x}S_{l+3}^{x}$, which have 
respectively projection orders
$m=1$, $2$, $3$ and $4$, as it can be easily computed 
identifying the lowest non-zero integer such that $\langle \mathcal O H^m \rangle$ is not zero (see the discussion in Sec.~\ref{SubSec:Estimating:m}).

First, in Fig.~\ref{Fig-DETH} we show their diagonal matrix elements in the energy eigenbasis for spin chains up to $L=12$. 
Since the energy density of Hamiltonian~\eqref{eq:H_Ising} vanishes at infinite temperature, we are interested in the behavior of the profiles at energy density close to $\varepsilon = 0$. 
The four panels show compatibility with the scaling $\mathcal{O}(\varepsilon)-\mathcal{O}(\varepsilon=0)\sim \varepsilon^m$: as expected from Sec.~\ref{SubSec:Estimating:m}, the overlap technique gives information on the diagonal ETH function.

Second, in Fig.~\ref{Fig-AAT} we show the autocorrelation functions 
$
C(t) \equiv \la \mathcal{O}(t)\mathcal{O}\ra_{L,c}
$
for the operators above at different system sizes, where the novel notation $C(t)$ is only introduced for plotting simplicity. 
Using dynamical quantum typicality (DQT)~\cite{Gemmer09-QDT,Robin14-QDT}, we are able to consider system sizes up to $L = 18$, see App.~\ref{app:DT} for more details on the numerical method and some additional plots.
We observe a transient behavior at small times and an eventual saturation to a size-dependent value at larger times.
In Fig.~\ref{Fig-LT} we discuss the fact that the saturation value decays as $L^{-m}$, as we expect from the analysis in Sec.~\ref{sec:fin_size}.

In the intermediate regime, an algebraic relaxation compatible with $t^{-m/2}$ is clearly seen.
Hence, we can conclude that for the four observables that we just considered, the relaxation-overlap inequality in Eq.~\eqref{eq:ineq} is not only satisfied but also saturated. We had anticipated this saturation in our hydrodynamic discussion, and specifically in the analytical results in Eq.~\eqref{eq:C_decay} of Sec.~\ref{sec:Diffusion}, obtained under the assumptions that (i) energy spreads diffusively through the system, and (ii) the hydrodynamic correlator $\tilde C(\mathbf k=0, t=0) \neq 0$.

Finally, we are ready to establish numerically the main result of the article.
We study the off-diagonal matrix elements and compute the off-diagonal ETH function using two different techniques.
The former is based on the following definition, well-suited for numerical studies based on exact diagonalization (ED) \cite{Kurchan19-GETH}:
\begin{equation}\label{eq-f-num}
|f_{\mathcal{O}}(\omega)|^{2}\equiv\frac{1}{{\cal Z}_{\beta}}\sum_{i\neq j}e^{-\frac{\beta}{2}(E_{i}+E_{j})}|{\cal O}_{ij}|^{2}\delta_{\epsilon}(\omega-(E_{i}-E_{j}))
\end{equation}
 where $ {\cal Z}_{\beta}=\sum_{i}e^{-\beta E_{i}}$ and 
$\delta_\epsilon(x)$ is a coarse-grained delta function defined for some small $\epsilon$ \cite{epsilon} as
\begin{equation}
\delta_{\epsilon}(x)=\begin{cases}
1/\epsilon, & |x|\le\epsilon/2;\\
0, & |x|>\epsilon/2.
\end{cases}
\end{equation}
In the thermodynamics limit $L\rightarrow \infty$,
\begin{equation}\label{eq-f-ed}
|f_{\mathcal{O}}(\omega)|^{2}=|f_{{\cal O}}(\varepsilon(\beta),\omega)|^{2}.
\end{equation}
The second is an indirect calculation
through the Fourier transform of the autocorrelation function (at inverse temperature $\beta$)
\begin{equation}\label{eq-f-typ}
|f_{{\cal O}}(\omega)|^{2}=\frac{1}{2\pi}e^{\frac{1}{2}\beta\omega}\int e^{-i\omega t}\left(C(t)-C(\infty)\right)dt.
\end{equation}
This second technique allows us to reach much larger system sizes.

The data, shown in Fig.~\ref{Fig-FW}, are consistent with a collapse in the limit $L\rightarrow \infty$; as we anticipated, finite size effects are enhanced at small frequencies and should not be considered in our analysis. 
A divergence $|f_\mathcal{O}(\omega)|^2 \sim 1/\sqrt{\omega}$ is observed clearly for $m=1$, as predicted from the theory. A logarithmic divergence is expected at $m=2$, as discussed in Sec.~\ref{sec:off_diag}, and the data display the divergence above. 
In the other cases ($m=3,4$) $|f_{\mathcal O}(\omega)|^2$ remains finite, but a singularity $|f_{\mathcal O}(\omega)|^2-|f_{\mathcal{O}}(0)|^2 \sim |\omega|^{m/2-1}$ is observed.
In summary, our numerics on the non-integrable spin-1 Ising model features all the scalings anticipated in Table~\ref{tab:Summary} and validates the existence of a connection between the diagonal and off-diagonal functions of ETH.

\begin{figure}[t!]
	\includegraphics[width=1.0\linewidth]{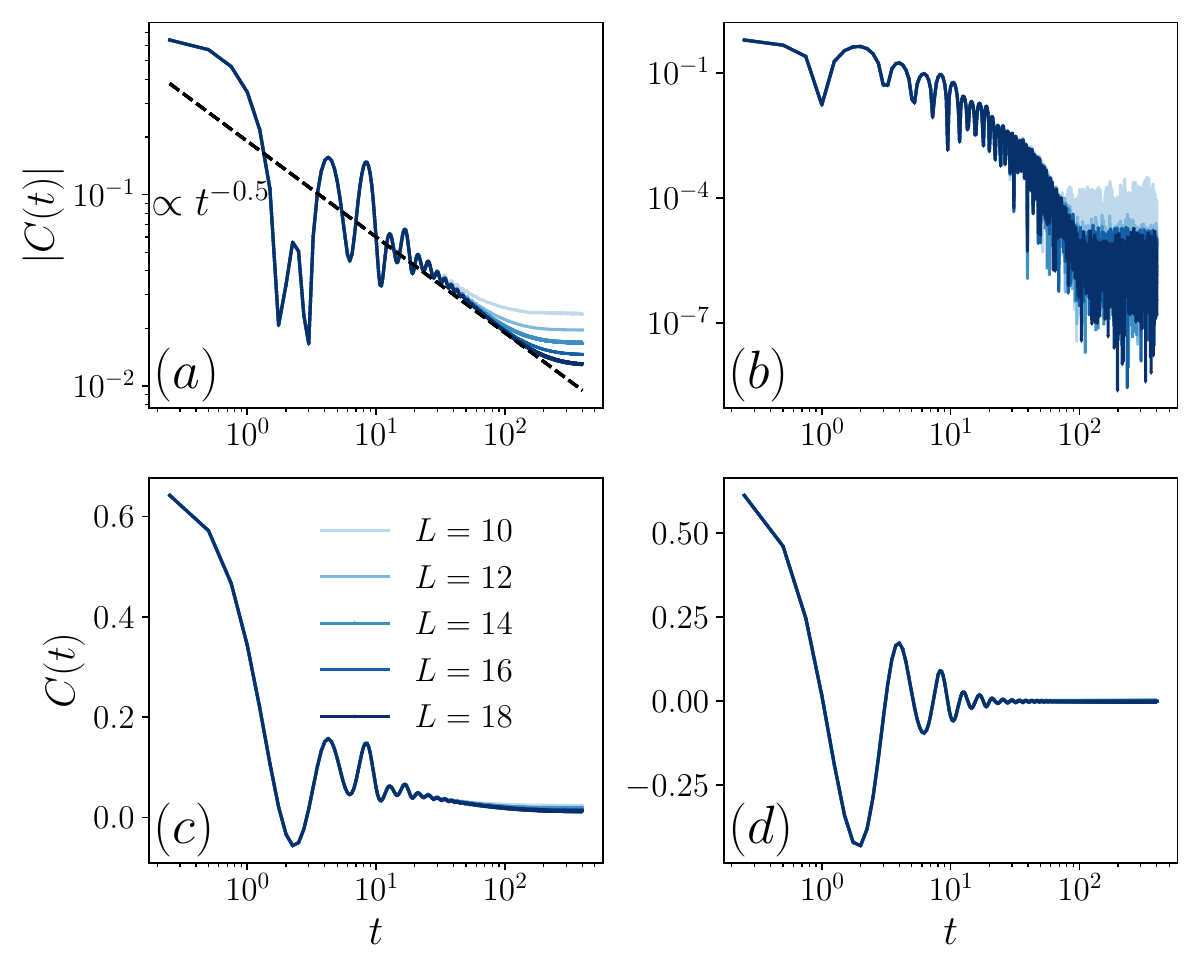}
 \caption{Autocorrelation function with/without absolute value ($|C(t)|,C(t)$ respectively) at infinite temperature in tilted field Ising model for the operators $S_{1}^{z}$ (panels (a) and (c)) and $S_{1}^{y}$ (panels (b) and (d)), for system sizes $L=10,12,14,16,18$. The dashed line indicates the expected leading scaling $\propto 1/t^{1/2}$. Slow decaying oscillations around $0$ are observed for the operator $S^y_1$. Results for $L=10$ are calculated by exact diagonalization, and for $L > 10$ results are calculated using DQT.
 }\label{Fig-AAT-YZ}
\end{figure}

\subsection{The case of a relaxation-overlap inequality that is not saturated}

We now present a numerical study of the mechanism discussed in Sec.~\ref{sec:t_der} that allows to systematically introduce operators whose autocorrelation functions satisfy but do not saturate the relaxation-overlap inequality~\eqref{eq:ineq}. 
The idea is based on taking operators that are the time derivative of a local operator; 
in our case, we can take the operator $S^y_l$, which satisfies
\be
S^y_l \propto i[H,S^z_l]
\ee
up to an irrelevant constant. 
In Fig.~\ref{Fig-AAT-YZ} we show the autocorrelation functions of $S^z_l$ and of $S^y_l$, computed numerically, as a function of time; the two different behaviours are compared and contrasted. 

Beginning with the operator $S^z_l$, we find a good agreement with the scaling $C(t)\sim 1/t^{1/2}$; 
additional oscillations are found, but their contribution is subdominant at later times. 
Since $S^z_l$ has overlap order $m=1$ (the Hamiltonian $H$ has a non-vanishing projection onto it), we conclude that for this operator the relaxation-overlap inequality is saturated. On the other hand, 
after computing the autocorrelation function for $S^y_l$, we verified numerically that it corresponds to the second time derivative of the autocorrelation function of $S^z_l$, as anticipated in Eq.~\eqref{eq:sec_derivative}.
Strong oscillations around the value $0$ are found for $S^y$, which hide the scaling $\sim 1/t^{5/2}$ that we expect from the aforementioned relation~\eqref{eq:sec_derivative} between the two autocorrelation functions: in particular, the algebraic decay above is not manifest in the panel (b) of Fig.~\ref{Fig-AAT-YZ}, which only highlights the decay of the oscillations (not predicted by our theory).
On the other hand, for the operator $S^y_l$, the overlap order is $m=\infty$, as it can be explicitly verified from the fact that $\langle S^y_l H^m \rangle =0$ for all $m$ or that the diagonal ETH function is identically equal to $0$ for any $\varepsilon$.

\begin{figure}[tb]
	\includegraphics[width=1.0\linewidth]{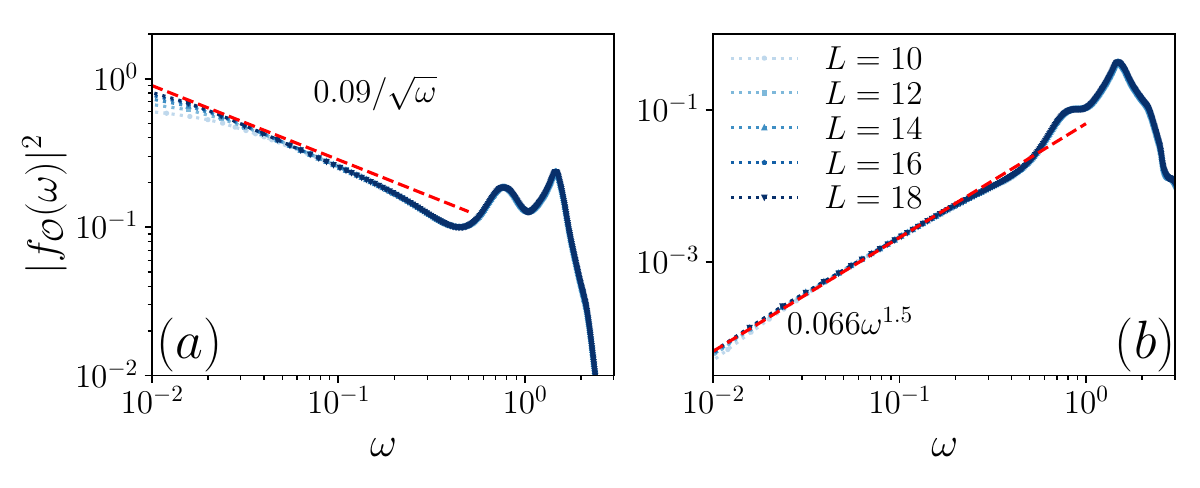}
 \caption{$|f_{\cal O}(\omega)|^2$ versus $\omega$ in tilted field Ising model, for the operators $S^z$ (a) and $S^y$ (b) for system size $L=10,12,14,16,18$ for $\beta = 0.0$. The dashed red line gives the singular behaviour ($\propto 1/\sqrt{\omega},\omega^{1.5}$ respectively) expected for small $\omega$. }\label{Fig-FW-YZ}
\end{figure}

To conclude, we probe the off-diagonal part of the ETH ansatz, which gives access to the frequency domain of the autocorrelation function, and we plot it in Fig.~\ref{Fig-FW-YZ}. We see that $|f_\mathcal{O}(\omega)|^2$ is compatible with a singular behavior $\propto 1/\sqrt{\omega}$ for the operator $S^z_l$, and $ \propto \omega^{3/2}$ for  $S_l^y$  in the limit $\omega \rightarrow 0$, as expected from the theory. 
Furthermore, additional peaks for other values of $\omega$ occur, which we relate to the slow decaying oscillations observed in real time. However, we do not have a theory to explain them, since pure diffusion, associated with monotonic decays of autocorrelation functions, does not predict the effects above. Finally, for the sake of completeness, we note that the behavior at small $\omega$ can be equivalently captured either by truncating high frequencies or by applying a moving average in the time domain, which effectively cancels oscillations: these two approaches are indeed connected through the Fourier transform.

\subsection{Long-range interactions and anomalous diffusion}

\begin{figure}[tb]
	\includegraphics[width=1.0\linewidth]{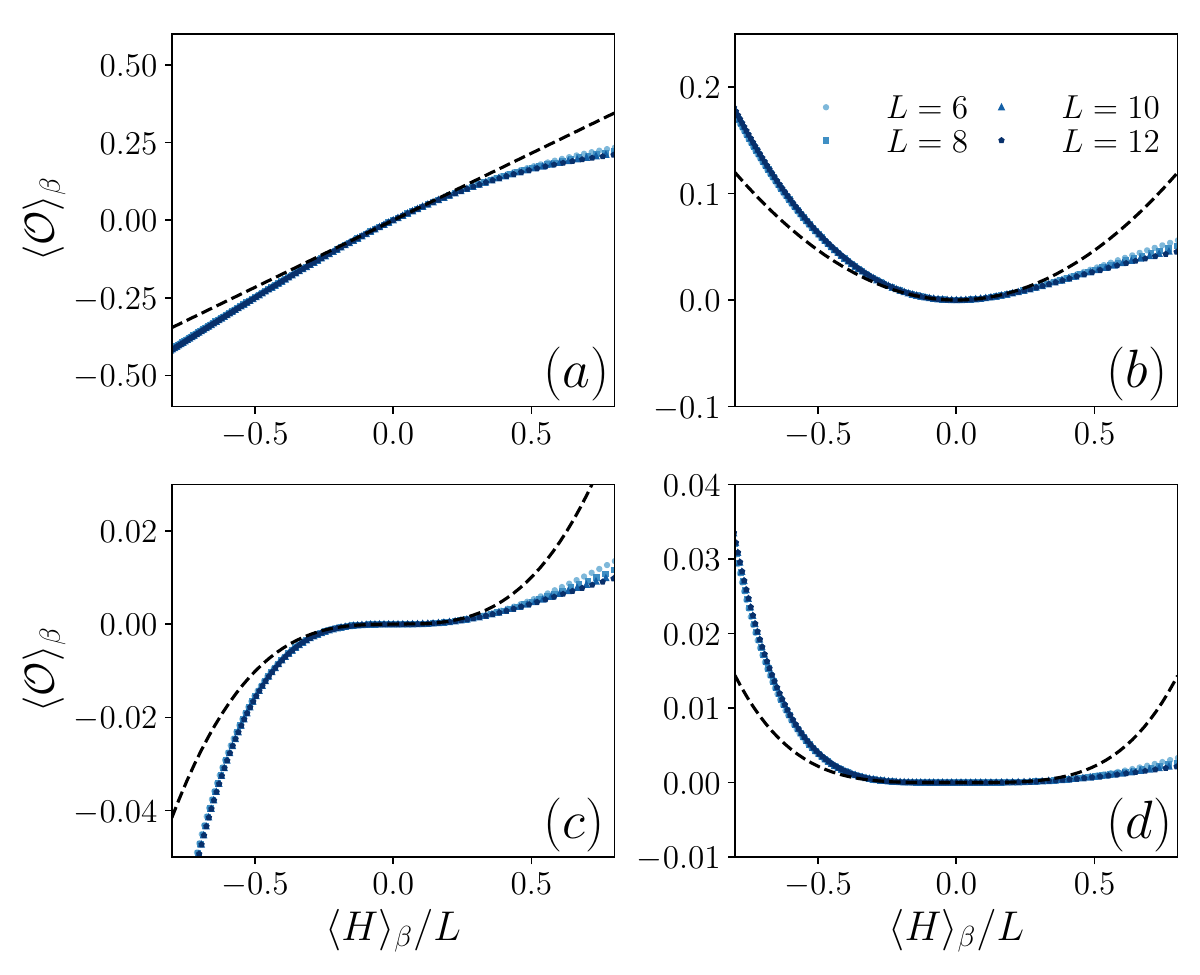}
 \caption{$\langle {\cal O}\rangle_\beta$ versus $\langle H \rangle_\beta/L$ in long-range Ising model, for (a) $S_{1}^{x}$ ; (b) $S_{1}^{x}S_{2}^{x}$; (c) $S_{1}^{x}S_{2}^{x}S_{3}^{x}$; (d) $S_{1}^{x}S_{2}^{x}S_{3}^{x}S_{4}^{x}$, for system size $L=6,8,10,12$. The dashed line shows the scaling  $\mathcal{O}(\varepsilon)\sim\varepsilon^{m}$ where $\varepsilon=\langle H\rangle_{\beta}/L$,  and the prefactor is calculated using Eq.~\eqref{eq:fit_formula}.}\label{Fig-DETH-LR}
\end{figure}

\begin{figure}[tb]
	\includegraphics[width=1.0\linewidth]{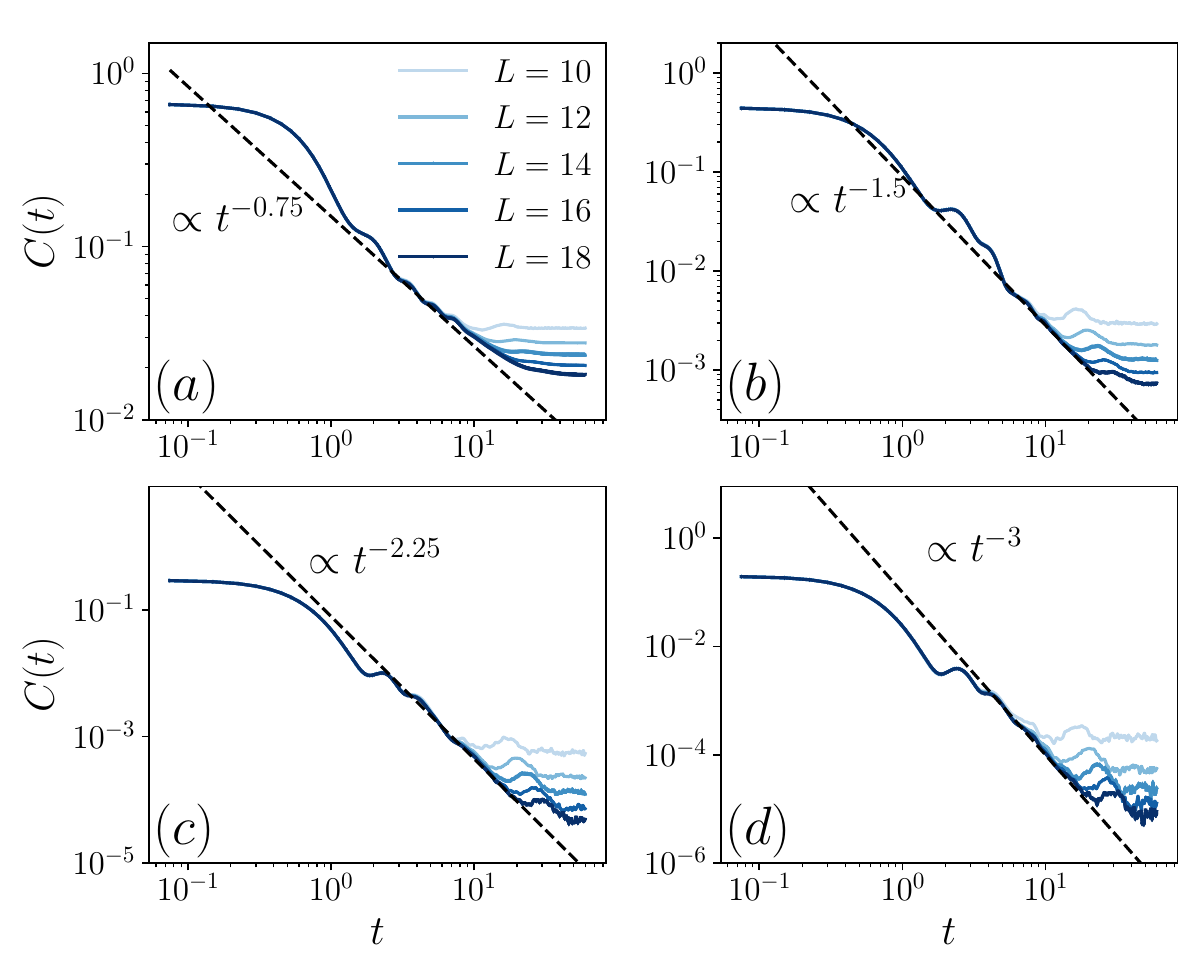}
 \caption{Autocorrelation function $C(t)$ at infinite temperature in long-range Ising model for the operators (a) $S_{1}^{x}$ ; (b) $S_{1}^{x}S_{2}^{x}$; (c) $S_{1}^{x}S_{2}^{x}S_{3}^{x}$; (d) $S_{1}^{x}S_{2}^{x}S_{3}^{x}S_{4}^{x}$, for system sizes $L=10, 12,14,16,18$. The dashed line indicates the expected scaling $\propto 1/t^{m/z}$, with $z = 4/3$. Results for $L=10$ are calculated by ED, and for $L > 10$ results are calculated using DQT.}\label{Fig-AAT-LR}
\end{figure}

\begin{figure}[tb]
\includegraphics[width=1.0\linewidth]{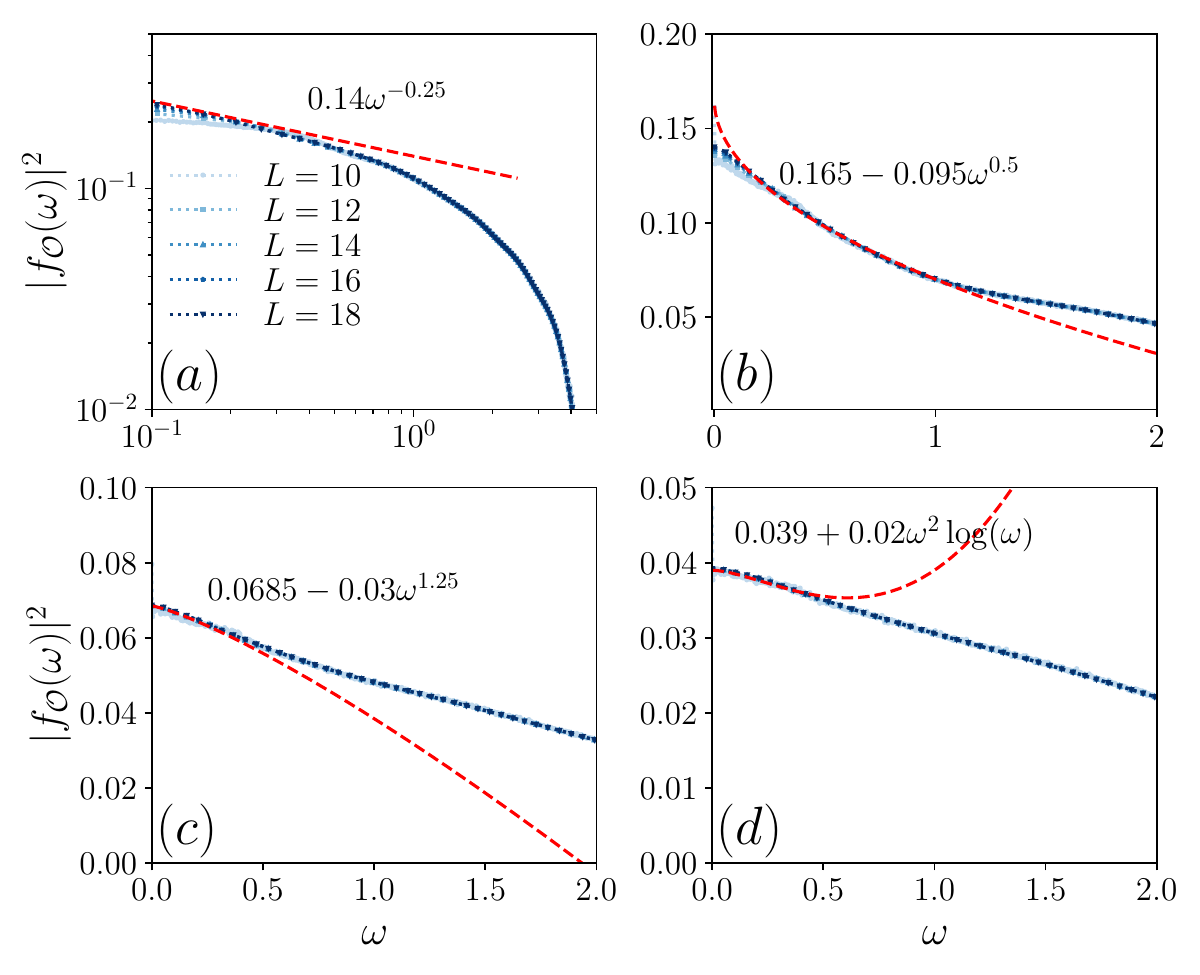}
 \caption{$|f_{\cal O}(\omega)|^2$ versus $\omega$, in long-range Ising model, for operators (a) $S_{1}^{x}$ ; (b) $S_{1}^{x}S_{2}^{x}$; (c) $S_{1}^{x}S_{2}^{x}S_{3}^{x}$; (d) $S_{1}^{x}S_{2}^{x}S_{3}^{x}S_{4}^{x}$, for system size $L=10,12,14,16,18$, for $\beta = 0.0$. Results for $L = 10$ are calculated by ED, while results for $L>10$ are calculated by the Fourier transform of the corresponding autocorrelation functions shown in Fig.~\ref{Fig-AAT-LR} computed using DQT.}\label{Fig-FW-LR}
\end{figure}

To check whether our main results can be generalized to systems exhibiting anomalous diffusion, 
we also consider a spin-$1$ Ising chain with transverse field and long-range coupling.
The Hamiltonian reads
\begin{equation}
H=\sum_{i=1}^{L}\sum_{j>i}\frac{J}{N_{\alpha}r_{ij}^{\alpha}}S_{i}^{z}S_{j}^{z}+\sum_{i=1}^{L}h_{x}S_{i}^{x},
\end{equation}
where 
\begin{equation}
    r_{ij}=\min(|j-i|,L-|j-i|),\ N_{\alpha}=\left(\sum_{i=2}^{L}\frac{1}{r_{i1}^{2\alpha}}\right)^{-\frac 12}.
\end{equation}
The parameters are chosen as $J = 2.0, h_x = 1.1, \alpha = 1.5$, where the system is non-integrable (in Appendix~\ref{app:DT} we discuss the level spacing statistics of the model) and is expected to exhibit super-diffusive transport~\cite{slk-20, Knap-22, ddmprt-23,Laurent-LR-Ising}. 
Similar to the short-range model, we now discuss results for diagonal matrix elements, autocorrelation functions, and off-diagonal matrix elements of the same four observables considered so far. 
Note that including long-range interactions does not change their overlap orders.

Since finite-size effects are stronger when dealing with power-law potentials, we prefer evaluating canonical expectation values instead of microcanonical ones. In Fig.~\ref{Fig-DETH-LR} we show $\langle {\cal O} \rangle_\beta$ as a function of $\langle H \rangle_\beta / L$, where $\langle\bullet\rangle_{\beta}:=\frac{\text{Tr}[e^{-\beta H}\bullet]}{\text{Tr}[e^{-\beta H}]}$ indicates the average over the thermal state. 
Agreement with
the scaling $\mathcal{O}(\varepsilon)-\mathcal{O}(\varepsilon=0)\sim \varepsilon^m$  (here $\varepsilon = \langle H \rangle_\beta/L$) is observed, which is expected from Sec.~\ref{sec:fin_size}. 
The value of $m$ corresponds exactly to the overlap order and
in particular, the black dashed line is computed using the exact formula in Eq.~\eqref{eq:fit_formula} with no fitting parameters.

The autocorrelation functions $C(t)$ are shown in Fig.~\ref{Fig-AAT-LR}. 
Before saturation, in the intermediate regime, an algebraic relaxation compatible with $t^{-dm/z}$ is observed with $d=1$ and $z\simeq 1.33$, which closely resembles the exponent $ z \approx 4/3$ found also in Ref.~\cite{Laurent-LR-Ising} for a long-range spin-1/2 model. 
The fact that the four decay exponents are proportional to $m$ indicates that the four observables saturate the general relaxation-overlap inequality in Eq.~\eqref{eq:gen_ROI}. 
More generally, these results
indicates that our predictions can also hold for super-diffusive systems, as argued in Sec.~\ref{sec:generaliz}.

In Fig.~\ref{Fig-FW-LR}, we show results of $|f_{\cal O}(\omega)|^2$ both via ED  and through the Fourier transform of autocorrelation, exactly as we did for the diffusive short-ranged model.
The  behavior $|f_{\cal O} (\omega)|^2\sim \omega^{-1/4}$ predicted by the theory is observed clearly for $m = 1$. For $m = 2$, the predicted behavior $|f_{\cal O}(\omega)|^2 - |f_{\cal O}(0)|^2\sim|\omega|^{1/2}$ can also be seen. For $m=3,4$ the curves are also compatible with our theoretical expectations, although finite-size effects are more pronounced and larger systems should be studied. 

Concluding, the numerical evaluations consistently corroborate our theoretical predictions concerning the functional dependence of the diagonal elements on energy, the singular behavior of the off-diagonal ETH function, and the long-time relaxation dynamics of two-time correlators.

\subsection{Analysis at finite temperature}\label{sec:finiteT} 
\begin{figure}[tb]
	\includegraphics[width=1.0\linewidth]{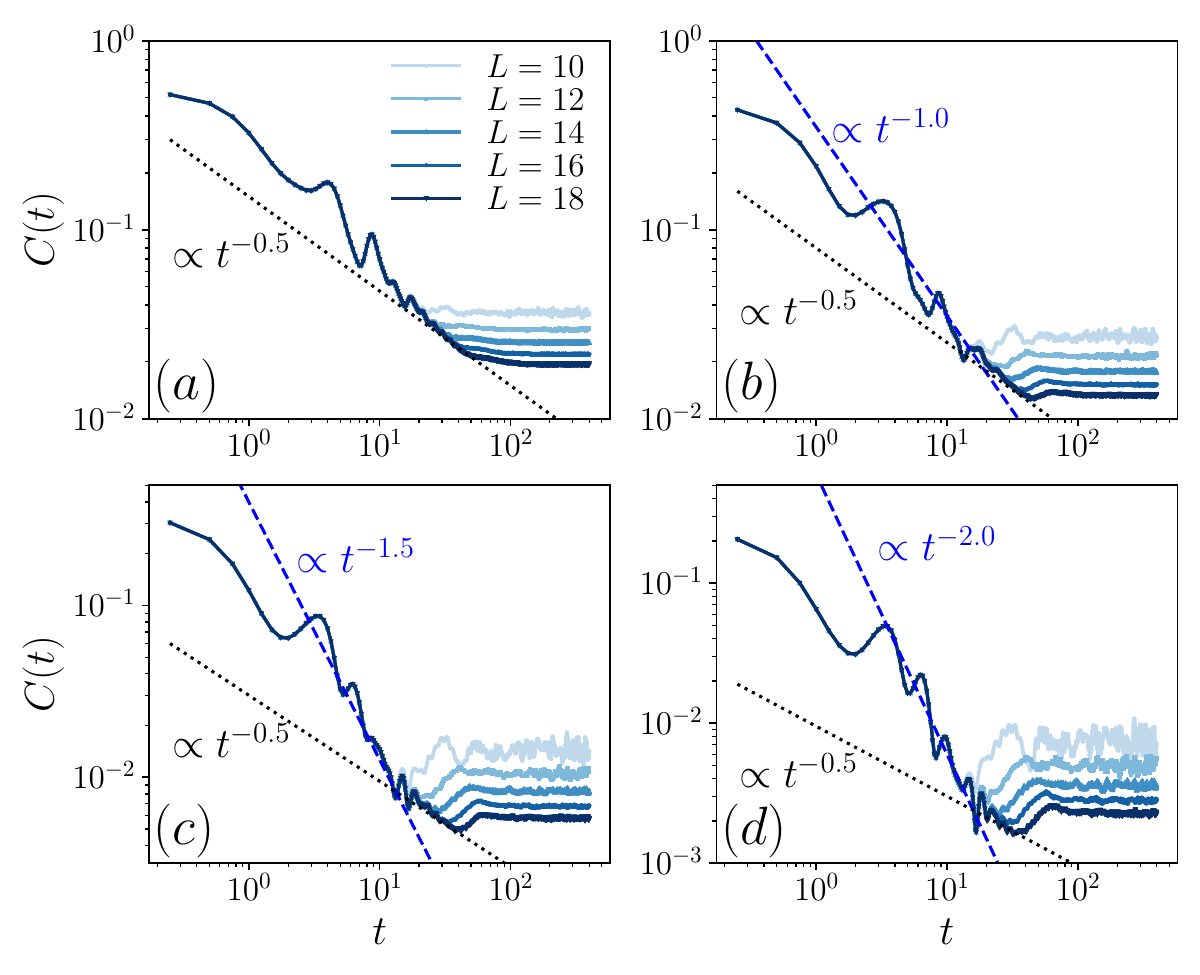}
 \caption{Autocorrelation function $C(t)$ at finite temperature $\beta = 0.6$ in the tilted field spin-1 Ising model for the operators (a) $S_{1}^{x}$; (b) $S_{1}^{x}S_{2}^{x}$; (c) $S_{1}^{x}S_{2}^{x}S_{3}^{x}$; (d) $S_{1}^{x}S_{2}^{x}S_{3}^{x}S_{4}^{x}$, for system sizes $L=10, 12,14,16,18$. The dashed and dotted lines indicate the scaling $\propto 1/t^{m/2}$ and $\propto 1/t^{1/2}$, respectively. Results for $L=10$ are calculated by exact diagonalization, and for $L > 10$ results are calculated using DQT. Here we only show the real part of $C(t)$.}\label{Fig-AAT-B}
\end{figure}
\begin{figure}[tb]
\includegraphics[width=1.0\linewidth]{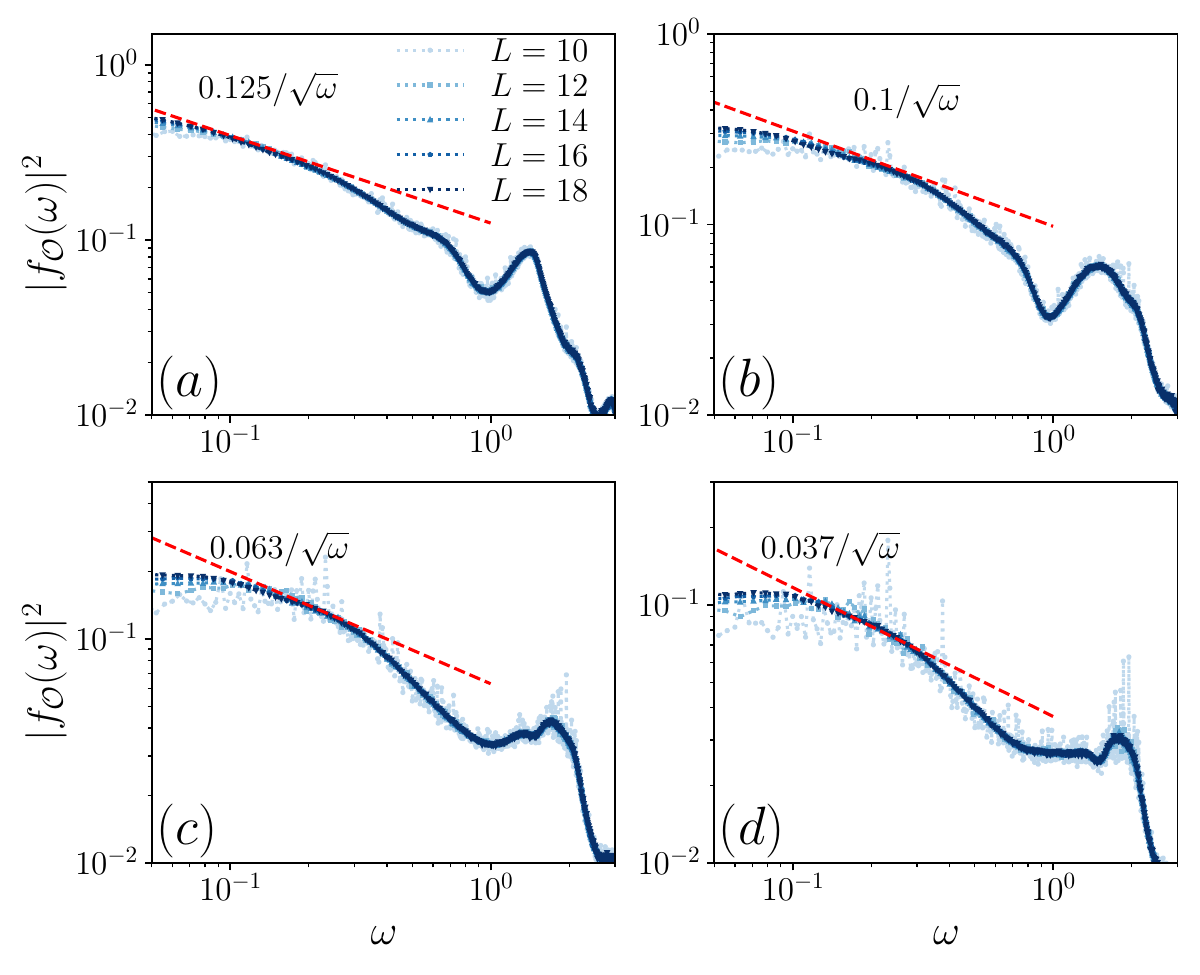}
 \caption{$|f_{\cal O}(\omega)|^2$ versus $\omega$, in the tilted field spin-1 Ising model for operators (a) $S_{1}^{x}$; (b) $S_{1}^{x}S_{2}^{x}$; (c) $S_{1}^{x}S_{2}^{x}S_{3}^{x}$; (d) $S_{1}^{x}S_{2}^{x}S_{3}^{x}S_{4}^{x}$, for system size $L=10,12,14,16,18$, for finite temperature $\beta = 0.6$. Results for $L = 10$ are calculated by ED and Eq.~\eqref{eq-f-num}, while results for $L>10$ are calculated using Eq.~\eqref{eq-f-typ}, with the corresponding autocorrelation functions shown in Fig.~\ref{Fig-AAT-B} by DQT. The dashed line gives the singular behavior $\propto 1/\sqrt{\omega}$.}\label{Fig-FW-Beta}
\end{figure}
In Fig. \ref{Fig-AAT-B} we study the finite temperature autocorrelation functions of the short-range Ising model, and we consider the same operators of Fig. \ref{Fig-AAT}. We choose $\beta=0.6$, where the overlap orders of these operators, identified by the Taylor expansion of $\mathcal{O}(\varepsilon)$ around $\varepsilon \simeq \varepsilon_\beta$ is $1$ (see also Fig. \ref{Fig-DETH}). Consequently, our theory predicts an eventual algebraic decay $\sim t^{-1/2}$ in the infinite volume limit. While this is observed at later times, before the saturation due to finite-size effects takes place, at shorter times the algebraic decay resembles that at the infinite temperature of Fig. \ref{Fig-AAT}. We understand this as transient behavior associated with a small $\beta$ expansion, whereby a crossover between the two aforementioned regimes is expected; similarly, given the eventual emergence of relaxation as $\sim t^{-1/2}$, we also expect that off-diagonal part will show a behavior as $|f_{\mathcal{O}}(\omega)|^2\sim 1/\sqrt{\omega}$, which we can observe in Fig.~\ref{Fig-FW-Beta}. A deeper investigation of the crossover time is beyond the purpose of this work and will pursued in future investigations. 

As a final remark, we point out that algebraic decays different from $\sim t^{-1/2}$ are in principle possible for a distinct choice of the local operators (compared with those in Fig.~\ref{Fig-AAT-B}). For example, $\mathcal{O} = \lambda_1 S^x_1 + \lambda_2 S^x_1S^x_2$ ($\lambda_1,\lambda_2>0$) has a microcanonical expectation value whose minimum is at energy density $\varepsilon_0 \neq 0$, as it can be deduced from the profiles of Fig.~\ref{Fig-DETH}; thus, from our theory, we would expect that the decay of its autocorrelator should go as $\sim t^{-1}$ when evaluated at inverse temperature $\beta\neq 0$ satisfying $\varepsilon(\beta) = \varepsilon_0$. Furthermore, by taking a proper fine-tuned linear combination of the operators considered in Fig.~\ref{Fig-DETH}, one can easily reconstruct an operator satisfying \eqref{eq:non_zero_der}, with $m=1,\dots,4$ at $\beta \neq 0$, and, therefore, with an expected associated decay as $\sim t^{-m/2}$.

\section{Conclusion}\label{sec:conclusions}

In this article we have discussed a relation between the diagonal and off-diagonal functions of the ETH ansatz, $\mathcal O(\varepsilon)$ and $f_{\mathcal O}(\varepsilon, \omega)$, respectively. In particular, for a given energy density $\varepsilon$, we have predicted the appearance of a non-smooth behavior in $|f_{\mathcal O}(\varepsilon, \omega)|^2$ at small frequencies that is related to the Taylor expansion of the diagonal ETH function. 
In the most common scenario, where $\mathcal O (\varepsilon') \simeq \mathcal O (\varepsilon) + \mathcal O'(\varepsilon) \times (\varepsilon' - \varepsilon)$ has a linear behaviour around $\varepsilon$ and energy spreads diffusively, a divergence $|f_{\mathcal{O}}(\varepsilon, \omega)|^2\sim 1/\sqrt{\omega}$ is expected and numerically observed. However, we have shown that an ample variety of singularities can occur for $\omega\rightarrow 0$, depending on the behaviour of the diagonal ETH function in $\varepsilon$. 
We have tested these predictions on a prototypical non-integrable spin model and shown agreement with numerical data for four different types of singularities; 
the regularization that we systematically observe at $\omega=0$ is a finite-size effect that is unavoidable even with our state-of-the-art algorithms. 

The discovery of this connection, as well as of the more general relaxation-overlap inequality, has been made possible by the use of a hydrodynamic approach to the autocorrelation function  $\langle \mathcal O(t) \mathcal O \rangle_c$.
Under the most common assumption, namely that the model supports the diffusive spreading of energy, which is also the unique local conserved quantity, 
it is possible to derive a set of quantitative relations for the behaviours of the ETH functions and of the autocorrelators associated to an operator $\mathcal O$ that are summarized in Tables~\ref{tab:Summary} and~\ref{tab:nu}. In fact, with the help of hydrodynamic assumptions only, it is possible to derive a generic relaxation-overlap inequality, linking the projection of an operator on the Hamiltonian and the time-decay of the autocorrelation function. 

While our discussion mostly refers to generic diffusive systems in one dimension at infinite temperature, we have also shown how this generalizes to the case of energy spreading anomalously due, for instance, to long-range couplings and to finite temperature.
We have presented a numerical analysis that verifies our predictions and shows how an anomalous dynamical exponent $z \neq 2$ affects the behaviour of ETH functions.
In general, our work shows that using hydrodynamic arguments can help unveil an underlying structure of
ETH has gone unnoticed so far.

\subsection{Perspectives}

We plan to expand our analysis along several different directions in order to explore further aspects of the connection between ETH and hydrodynamics.

The presence of additional conserved quantities, as for instance the magnetization in a XY spin chain with short- or long-range couplings~\cite{lp-20,lpfap-23,dlp-24}, can be taken into account via a slight generalization of our framework, including projections to all conserved charges. 
The phenomenology might be extremely rich and various since they may have a different dynamical exponent $z$.
Moving towards integrable models, where the set of integrals of motion is extensive, 
it would be interesting to investigate a possible connection between the well-established generalized hydrodynamics and the ETH for integrable models that has recently started to attract an increasing amount of interest~\cite{Alba-15,mv-20,ed-23}.

As a second point, we remark that in this article we have only focused on autocorrelation functions that decay algebraically in time.
We point out that while an exponential decay is usually associated with the absence of energy conservation, as for generic quantum circuits where the evolution is time-discrete~\cite{nrvh-17,nvh-18,vrps-18,fp-21}, exceptions are present. 
For example, conformal field theories are known to show an exponential time decay at finite temperatures despite supporting perfect ballistic transport~\cite{dkm-19}. 
We believe that such unusual behavior is ultimately related to the presence of peculiar strong revivals at finite sizes of autocorrelation functions~\cite{Cardy-14}, which makes impossible the use of our theory, which assumes the monotonic decay. 
This point deserves further investigation.

We also note that, in principle, singularities could occur at other values of the frequency $\omega$ besides $\omega = 0$, giving rise to slow decaying oscillations that mask the pure hydrodynamic and algebraic relaxation. 
We believe that the behaviour of the $ S_l^y$ observable discussed in the context of the short-range model, with strong oscillations in time of the autocorrelation function (see Fig.~\ref{Fig-AAT-YZ}), might be due to such singularities.
At the moment we do not have a theory for predicting such a mechanism, since hydrodynamics is not predictive for those effects.
Developing one such theoretical framework, capable of describing such oscillations, is of extreme interest.

Additionally, we would like to investigate systematically and rigorously the theory of hydrodynamic projections proposed in Sec.~\ref{sec:Diffusion} and discussed more extensively in Appendix~\ref{app:proj}. 
In the literature, we can find several
prototypical models where diffusive behaviours are established rigorously from first principles, as in quantum circuits with conserved charges~\cite{rpv-18,kvh-18} or free fermions with noisy hopping~\cite{bt-19}. 
It is imperative to test the theory that we presented in this article in such models.

This article has put the emphasis onto the two ETH functions, $\mathcal O(\varepsilon)$ and $f_{\mathcal O}(\varepsilon, \omega)$.
The possibility of reconstructing them efficiently for large system sizes has been recently pointed out in the context of energy filtering using a non-local non-Hermitian time evolution~\cite{bhc-20,ltbc-23,mcfm-24}.
It is to investigate whether such approach could provide valuable insights into the relations between the two ETH functions.

Recently, it has also been shown that the presence of steady non-equilibrium transport between two baths can be related to the autocorrelation functions and thus to the off-diagonal part of ETH for the current operator~\cite{xgp-2022,xgp-2023}. 
Our work suggests that it could be possible to interpret such a mechanism through the lens of diagonal ETH.

Finally, our predictions on the diffusive and anomalous relaxation-overlap inequality could be tested in experimental setups such as Rydberg atoms \cite{Bernien-2017, bl-2020}, superconducting analogue quantum processors \cite{ZhangPoletti2024} and trapped ions \cite{Joshi-2020}. The latter would be particularly suitable for this purpose, as it can readily produce both short- and long-range couplings, and it allows to explore both normal and anomalous diffusion. 

\acknowledgements

We acknowledge fruitful discussions with A.~De Luca and warmly thank M.~Fagotti for critical comments and extended discussions in the early stages of the project. 
D.P.~is supported from the Ministry of Education Singapore, under Grant No. MOE-T2EP50120-0019. X.X.~acknowledges support from joint Israel-Singapore NRF-ISF Research Grant No.~NRF2020-NRF-ISF004-3528.  
J.W.~acknowledges support from
the Deutsche
Forschungsgemeinschaft (DFG), under Grant No. 531128043, as well as under Grant
No.\ 397107022, No.\ 397067869, and No.\ 397082825 within the DFG Research
Unit FOR 2692, under Grant No.\ 355031190.
This work is part of HQI (www.hqi.fr) initiative and is supported by France 2030 under the French National Research Agency grant number ANR-22-PNCQ-0002;
L.M.~also acknowledges funding from the ANR project LOQUST ANR-23-CE47-0006-02. L.C.~acknowledges support from ERC Starting grant 805252 LoCoMacro.
Additionally, we greatly acknowledge computing time on the HPC3 at the University of Osnabr\"{u}ck, granted by the DFG, under Grant No. 456666331.

\bibliography{bibliography}

\begin{thebibliography}{109}
\bibitem [{eps()}]{epsilon}%
  \BibitemOpen
  \href@noop {} {}\bibinfo {note} {We choose $\epsilon = 0.004$ in our numerical simulations.}\BibitemShut {Stop}%
\end{thebibliography}

\clearpage
\newpage

\setcounter{figure}{0}     
\renewcommand{\thefigure}{A\arabic{figure}}

\begin{appendix}

\section{Analytical predictions for $\la \bar{\mathcal{O}}^2\ra_{L,c}$ in Eq.~\eqref{Eq:Fluctuations:OBar}}\label{app:Anal_pred}

In this Appendix, we give an analytical prediction for the finite-size fluctuations of the operator $\bar{\mathcal{O}}$, and we compute the proportionality constant in Eq.~\eqref{Eq:Fluctuations:OBar}.

First, we express the (central) moments of the Hamiltonian in Eq.~\eqref{Eq:Hamiltonian:moment:m} using the properties of Gaussian variables as
\be
\begin{split}
\la (H-&L\varepsilon(\beta))^m\ra_L \simeq \\
&\simeq\la (H-L\varepsilon(\beta))^2\ra^{m/2}_L \times \begin{cases} (m-1)!! \text{ for } m \text{ even},\\
0 \text{ for } m \text{ odd}.
\end{cases}
\end{split}
\ee
While the result above holds exactly if $H$ is a Gaussian variable, it gives in general the most leading term of the even moments, proportional to $\sim L^{m/2}$, as long as the central limit theorem holds. Also, even if the odd central moments might not be exactly vanishing, they grow slower than $L^{m/2}$, and therefore they can be neglected in the forthcoming computation.

Then, we compute $\la \bar{\mathcal{O}}^2\ra_{L,c}$ in Eq. \eqref{Eq:Fluctuations:OBar} and we obtain
\be
\begin{split}
\la \bar{\mathcal{O}}^2\ra_{L,c} \simeq& \l\frac{1}{m!}\frac{d^m\mathcal{O}}{d\varepsilon^m}\r^2 \times \\
& \times \l \la(H/L-\varepsilon(\beta))^{2m}\ra_L - \la(H/L-\varepsilon(\beta))^{m}\ra_L^{2}\r \simeq\\
\simeq &\l\frac{1}{m!}\frac{d^m\mathcal{O}}{d\varepsilon^m}\r^2 \times \la (H/L-\varepsilon(\beta))^2\ra^{m} \times\\
&\times \begin{cases} (2m-1)!!-\l(m-1)!!\r^2 \text{ for } m \text{ even},\\
(2m-1)!! \text{ for } m \text{ odd}.
\end{cases}
\end{split}
\ee
In particular, since $ \la(H/L-\varepsilon(\beta))^{2}\ra_L\sim 1/L$, the scaling $\la \bar{\mathcal{O}}^2\ra_{L,c}\sim L^{-m}$ holds for $m$ both even or odd.

\section{Hydrodynamic projections}\label{app:proj}

In this Appendix, we provide a definition of the normal ordering that appears in Eq.~\eqref{eq:naive_proj}. 
The general idea relies on the construction of spaces of local operators generated by the powers of $\{h(x)\}_x$ at a given order, where any possible contribution coming from projections onto lower orders is explicitly subtracted. We employ a formalism that is common in the mathematical formulation of hydrodynamics, that is the notion of hydrodynamic projections and generalization thereof, and we refer the reader to Refs.~\cite{Doyon-20,Doyon-22,Doyon-22a} for additional details.

Let us consider the algebra of local operators $\mathcal{A}$ \cite{Bratteli-12}, that is a Hilbert space with the inner product
\be
(a,b) \equiv \la a^\dagger b \ra,
\ee
where $\la \dots\ra$ is a reference clustering state (say, the thermal state considered in the main text). $\mathcal{A}$ contains a trivial subspace generated by the identity, denoted here by
\be
V_0 \equiv  \text{Span}\{1\}.
\ee
The orthogonal complement $V^{\perp}_0$ contains the operators with vanishing expectation value, and the projection $\mathcal{A}\rightarrow V^{\perp}_0$ is simply expressed by $\mathcal{O}\rightarrow \mathcal{O}-\la \mathcal{O}\ra$. We construct $V_1$ as the vector space generated by $h(x)-\la h(x)\ra$ inserted at a generic point $x$, that is
\be
V_1 \equiv \underset{x}{\text{Span}} \{h(x)\}  \bigcap V^{\perp}_0.
\ee
We generalize this procedure, and we systematically define the space $V_n$ generated by $\{h(x_1)\dots h(x_n)\}$ with vanishing projection onto $V_{n'},n'<n$. Specifically, we construct
\be
V_n \equiv  \underset{x_1,\dots,x_n}{\text{Span}} \{h(x_1)\dots h(x_n)\}  \bigcap_{n'=0}^n V^{\perp}_{n'}.
\ee
We will call \textit{normal ordering} the projection of $h(x_1)\dots h(x_n)$ onto $V_n$, and we denote it by
\be
\normOrd{h(x_1)\dots h(x_n)} \in V_n
\ee

Finally, we define the algebra generated by the Hamiltonian density as
\be
\mathcal{V} \equiv \overline{\bigoplus^{\infty}_{n=0} V_n},
\ee
where $\overline{(\dots)}$ denotes the closure with respect to the inner product. The space $\mathcal{V}$ is, roughly speaking, the space of operators generated by the Hamiltonian density and powers thereof, and in general it is strictly contained in the space of local operators $\mathcal{V} \subset \mathcal{A}$. The (hydrodynamic) projection
\be
\mathcal{A} \rightarrow \mathcal{V},
\ee
allows to formalize the notion of expansion of any local operator $\mathcal{O}(x)$ as a function of the Hamiltonian density appearing in Eq.~\eqref{eq:naive_proj}. In particular, the projection at order $n$, that is a map $\mathcal{A} \rightarrow V_n$, extracts from $\mathcal{O}(x)$ a contribution that is a linear combination of the operators $\normOrd{h(x_1)\dots h(x_n)}$. Also, the leading term comes from the points $x_1\simeq \dots \simeq x_n \simeq x$: for this reason, we formally express the contribution above as a Taylor series in $\normOrd{\partial^{p_1}_x h(x)\dots \partial^{p_n}_x h(x)}$, and we eventually obtain the expansion~\eqref{eq:naive_proj}.

\setcounter{figure}{0}
\renewcommand{\thefigure}{C\arabic{figure}}

\section{Numerical method and additional numerical results}\label{app:DT}

\begin{figure}[tb]
	\includegraphics[width=1.0\linewidth]{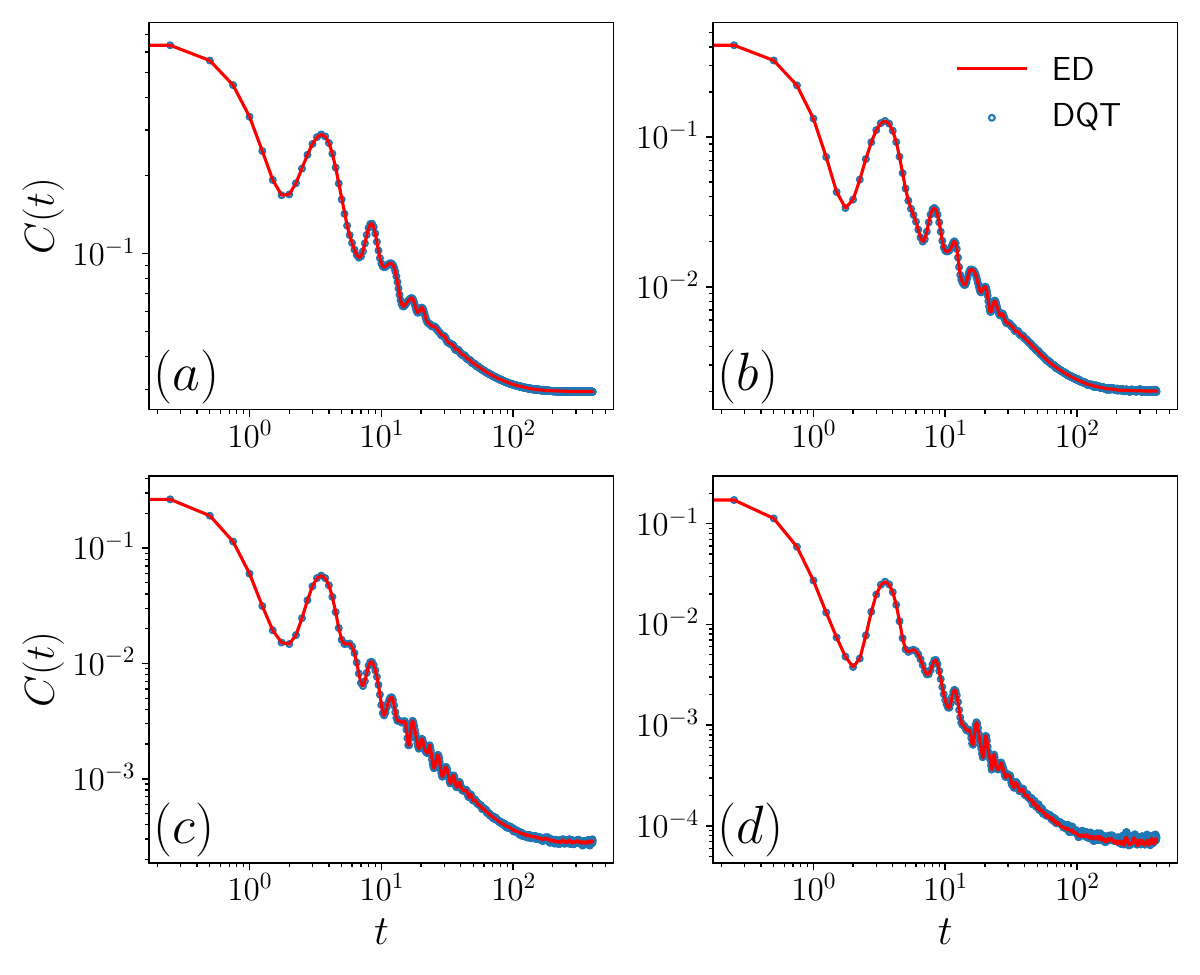}
 \caption{Comparison of results for autocorrelation function $C(t)$ in titled field Ising model at infinite temperature $\beta = 0.0$,  calculated by ED (red solid line) and DQT (blue open circle), for operators (a) $S_{1}^{x}$ ; (b) $S_{1}^{x}S_{2}^{x}$; (c) $S_{1}^{x}S_{2}^{x}S_{3}^{x}$; (d) $S_{1}^{x}S_{2}^{x}S_{3}^{x}S_{4}^{x}$. The system size $L=10$.}\label{Fig-Check-AAT}
\end{figure}

\begin{figure}[tb]
	\includegraphics[width=1.0\linewidth]{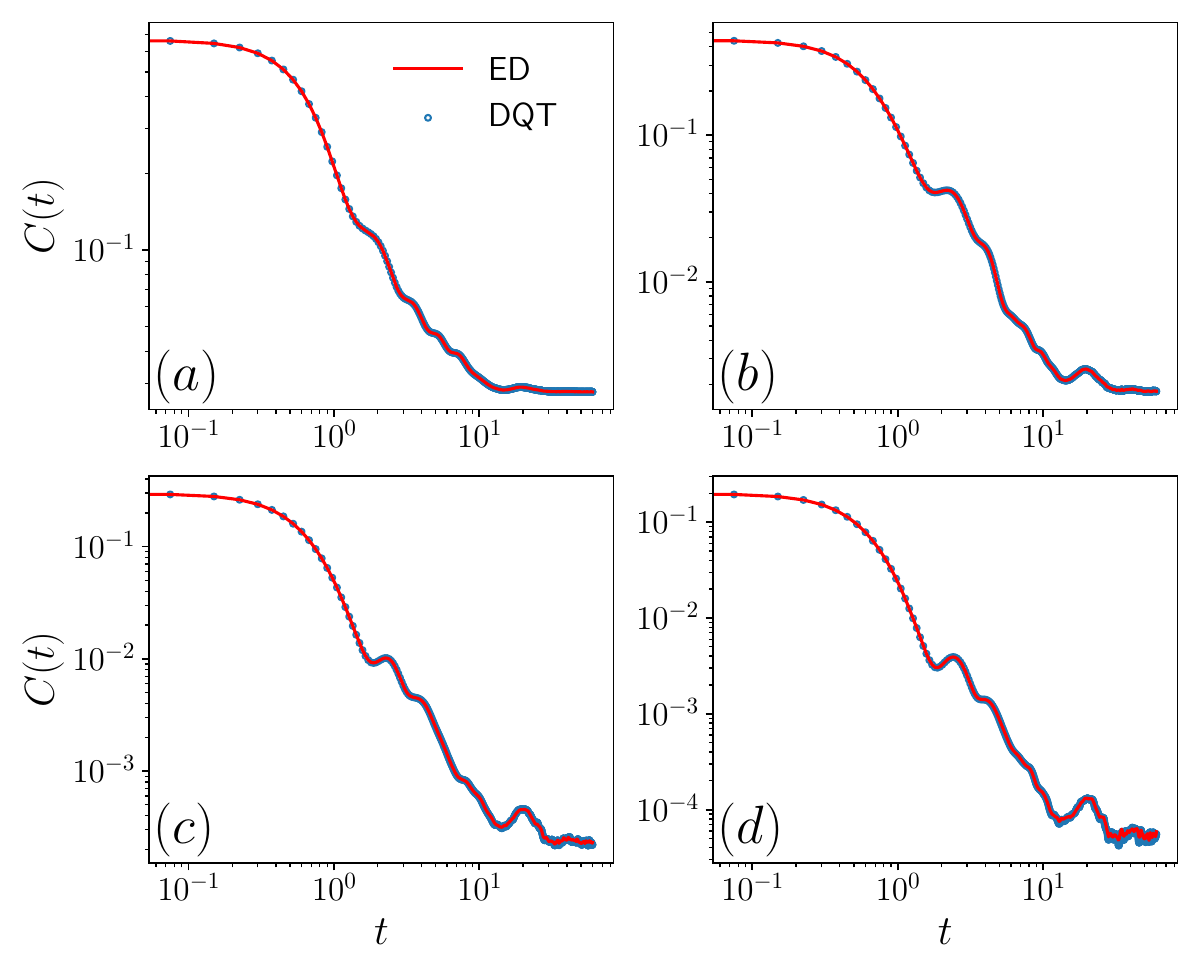}
 \caption{Similar to Fig.~\ref{Fig-Check-AAT}, but for long-range Ising model. }\label{Fig-Check-AAT-LR}
\end{figure}

In this Appendix, we briefly introduce the numerical method used in the main text to calculate autocorrelation functions, which is based on the dynamical quantum typicality (DQT) \cite{Robin14-QDT,Gemmer09-QDT}, and compare the results to exact diagonalization (ED) results.

\begin{figure}[tb]
	\includegraphics[width=1.0\linewidth]{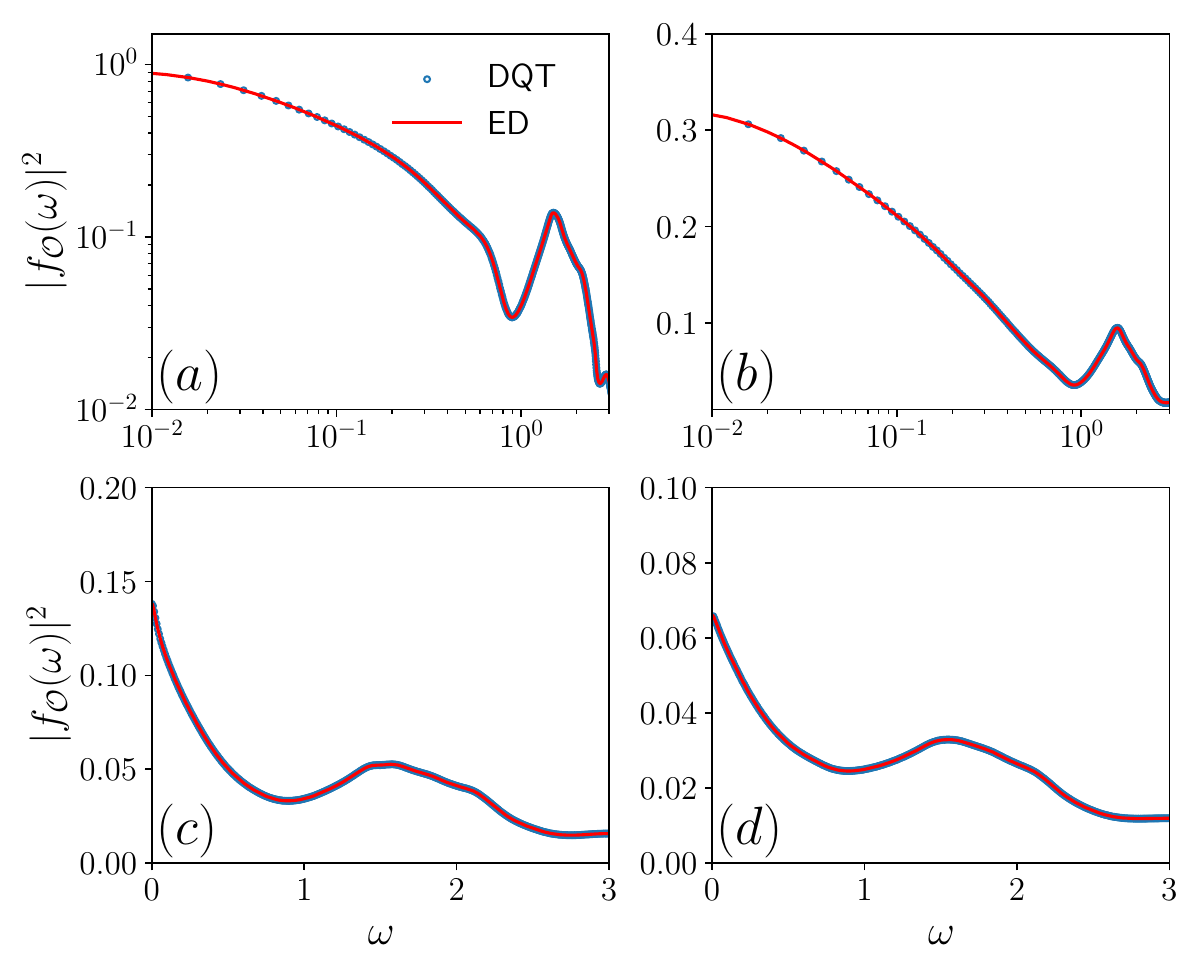}
 \caption{Comparison of results for $|f_{\mathcal O} (\omega)|^2$ in tilted field Ising model, for $\beta = 0.0$.  The red solid line indicates results calculated by ED, and the blue open circle indicates results calculated by the Fourier transform of the autocorrelation function (blue open circle in Fig. \ref{Fig-Check-AAT}) obtained by DQT.
The operators are (a) $S_{1}^{x}$ ; (b) $S_{1}^{x}S_{2}^{x}$; (c) $S_{1}^{x}S_{2}^{x}S_{3}^{x}$; (d) $S_{1}^{x}S_{2}^{x}S_{3}^{x}S_{4}^{x}$ and system size $L=12$.}\label{Fig-Check-FW}
\end{figure}

\begin{figure}[tb]
	\includegraphics[width=1.0\linewidth]{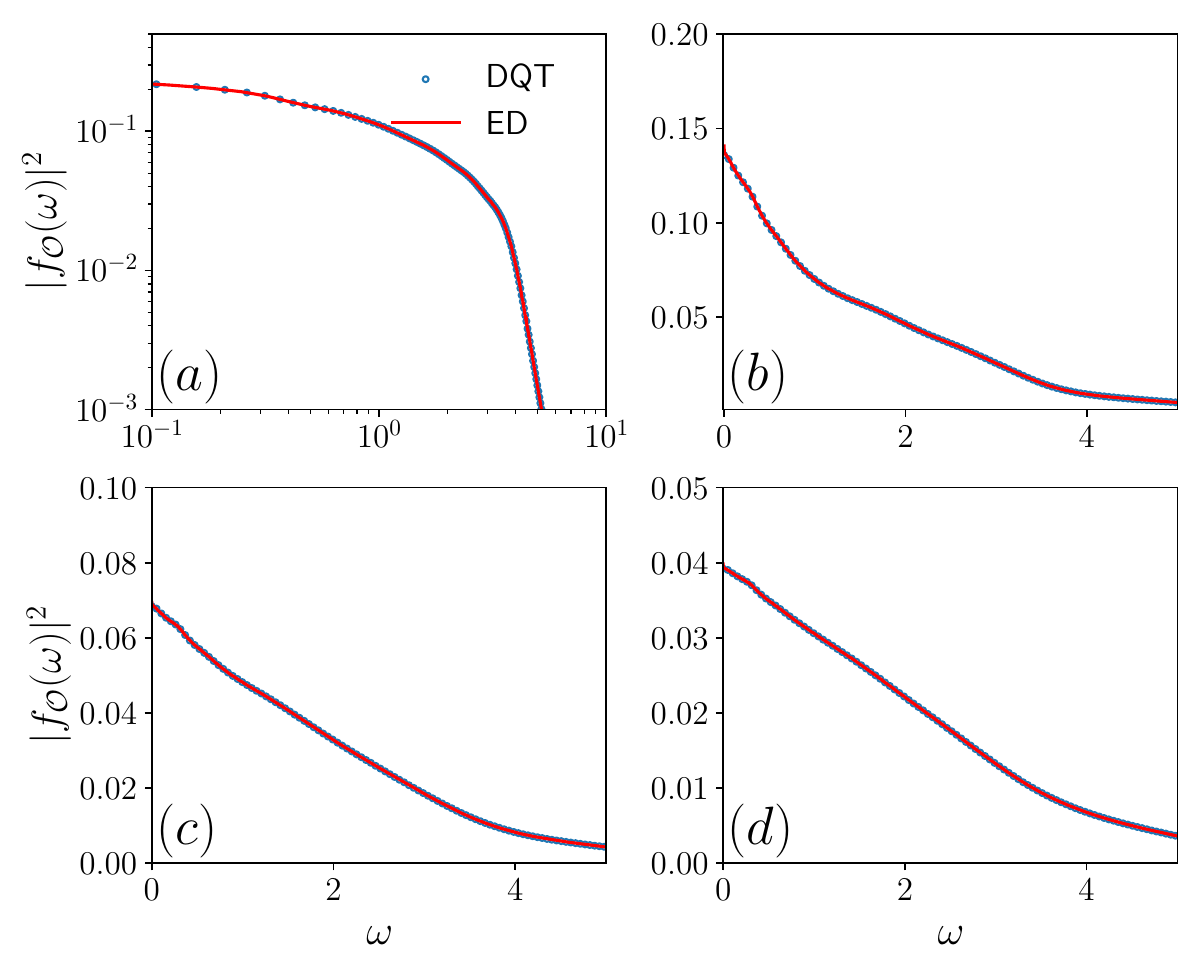}
 \caption{Similar to Fig.~\ref{Fig-Check-FW}, but for long-range Ising model.}\label{Fig-Check-FW-LR}
\end{figure}

Let us start from a normalized Haar random state in the Hilbert space
\begin{equation}
    |\psi\rangle = \sum_j C_j |E_j\rangle,
\end{equation}
where $C_j$ are complex numbers whose real and imaginary parts are drawn independently from a Gaussian distribution. 
According to DQT, the autocorrelation function at inverse temperature $\beta$, $C_\beta(t)=\frac{1}{Z_{\beta}}\text{Tr}[e^{-\beta H}{\cal O}(t){\cal O}]$ where $Z_{\beta}=\text{Tr}[e^{-\beta H}]$, can be written as
\begin{equation}
 C_{\beta}(t)=\frac{1}{\langle\psi|e^{-\beta H}|\psi\rangle}\langle\psi|e^{-\frac{\beta}{2}H}e^{iHt}{\cal O}e^{-iHt}{\cal O}e^{-\frac{\beta}{2}H}|\psi\rangle+\varepsilon(t).
\end{equation}
The error scales as
\begin{equation}
    \varepsilon(t)\sim\frac{1}{\sqrt{D}}
\end{equation}
with $D$ being the Hilbert space dimension of the system.
If $D$ is sufficiently large, one has
\begin{equation}\label{eq-Ct-typ}
C_{\beta}(t)\simeq\frac{1}{\langle\psi|e^{-\beta H}|\psi\rangle}\langle\psi|e^{-\frac{\beta}{2}H}e^{iHt}{\cal O}e^{-iHt}{\cal O}e^{-\frac{\beta}{2}H}|\psi\rangle.
\end{equation}
The accuracy of Eq.~\eqref{eq-Ct-typ} can be further improved by taking an average over $N_p$ different realizations of Haar random state $|\psi\rangle$. In our simulations, we choose $N_p \sim p(L)3^{-L}$, with $p(L)$ a non-decreasing function of $L$. In this way, we make sure that the accuracy of results does not decrease with increasing system size $L$. 
Employing two auxiliary states
\begin{equation}
    |\psi_{\beta}\rangle=e^{-\frac{\beta}{2}H}|\psi\rangle,\quad|\psi_{\beta}^{{\cal O}}\rangle={\cal O}e^{-\frac{\beta}{2}H}|\psi\rangle,
\end{equation}
Eq. \eqref{eq-Ct-typ} can be written as
\begin{equation}
  C_{\beta}(t)=\frac{1}{\langle\psi_{\beta}|\psi_{\beta}\rangle}\langle\psi_{\beta}(t)|{\cal O}|\psi_{\beta}^{{\cal O}}(t)\rangle,
\end{equation}
where
\begin{equation}
    |\psi_{\beta}(t)\rangle=e^{-iHt}|\psi_{\beta}\rangle,\quad|\psi_{\beta}^{{\cal O}}(t)\rangle=e^{-iHt}|\psi_{\beta}^{{\cal O}}\rangle.
\end{equation}
The time evolution of the two states can be
calculated by
iterating in real-time  using, e.g., Runge-Kutta method \cite{Boris13-RK,Robin14-RK}
or Chebyshev polynomial algorithm \cite{Jin10-Chebyshev,Raedt07-Chebysev}.
With similar methods, $|\psi_\beta\rangle$ (as well as $|\psi_\beta^{\cal O}\rangle$) can be calculated by iterating in imaginary time.
In this paper, we use Chebyshev polynomial algorithm, which allows us to calculate the autocorrelation function up to Hilbert space dimension $D = 3^{18}\approx 400000000$, far beyond the limit of ED.

\begin{figure}[tb]
	\includegraphics[width=1\linewidth]{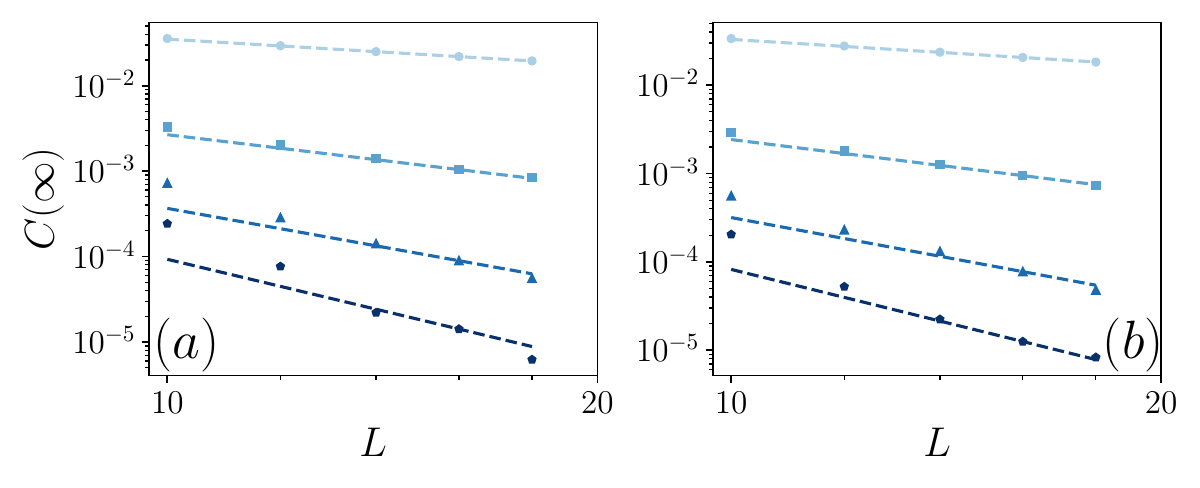}
 \caption{Saturation value of the autocorrelation function $C(\infty)$ versus system size, for (a)
titled field Ising model; (b) long-range Ising model. The operators considered are $S_{1}^{x},\ S_{1}^{x}S_{2}^{x},\ S_{1}^{x}S_{2}^{x}S_{3}^{x},\ S_{1}^{x}S_{2}^{x}S_{3}^{x}S_{4}^{x}$ (from light to dark blue). The dashed lines indicate the theoretical scaling $\sim 1/L^{m}$ (Eq.~\eqref{eq-Ldm}).
 }\label{Fig-LT}
\end{figure}

\begin{figure}[tb]
	\includegraphics[width=0.8\linewidth]{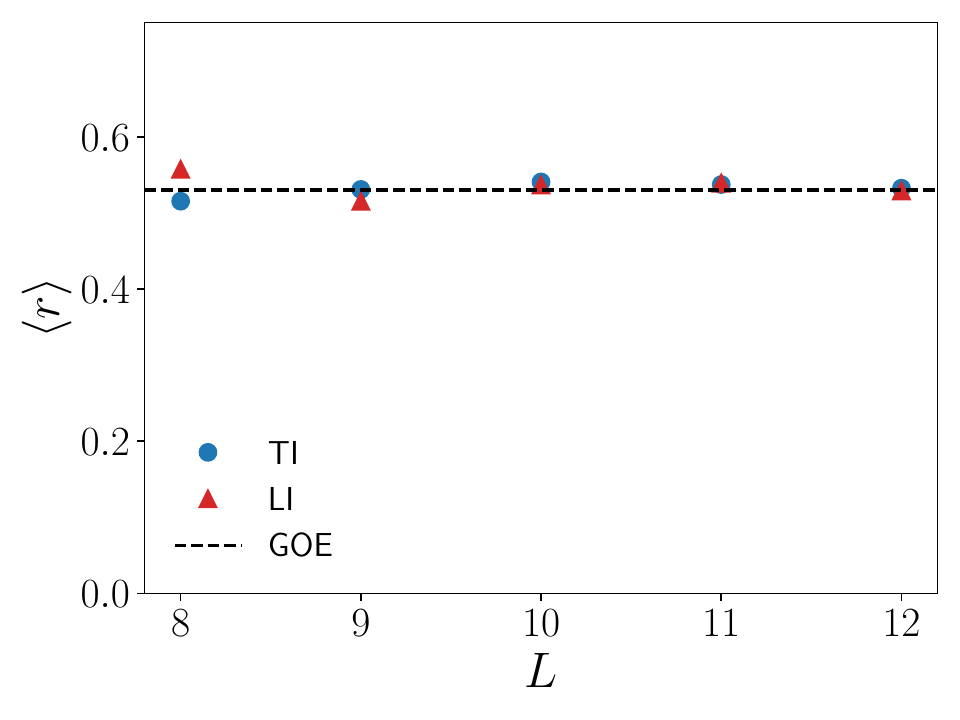}
 \caption{Mean adjacent gap ratio $\langle r \rangle$ versus system size $L$, in the tilted field Ising model (TI) and the long-range Ising model (LI). The dashed line indicates the prediction of GOE, $\langle r \rangle \approx 0.53$.}\label{Fig-Ar}
\end{figure}

To check the accuracy of our DQT results, we compare them to the results given by ED, for system size $L = 12$. For all operators considered, a perfect agreement between the DQT (with an additional averaging over different random states) and ED results can be found in Fig.~\ref{Fig-Check-AAT} and \ref{Fig-Check-AAT-LR}. 

Moreover, we also compare the results of $|f_{\mathcal O} (\omega)|^2$ given by the Fourier transform of the autocorrelation function (calculated using DQT) with the results from direct ED simulation. Perfect agreement is observed (Fig.~\ref{Fig-Check-FW} and \ref{Fig-Check-FW-LR}) for $L = 12$, at least for the region of $\omega$ we are interested in.

The results in Figs.~\ref{Fig-Check-AAT}, \ref{Fig-Check-AAT-LR}, \ref{Fig-Check-FW} and \ref{Fig-Check-FW-LR}, where have considered the same Hamiltonian parameters as in the main text, indicate the high accuracy of our DQT results, that are presented in Figs. \ref{Fig-AAT}, \ref{Fig-FW}, \ref{Fig-AAT-YZ}, \ref{Fig-FW-YZ}, \ref{Fig-AAT-LR} and \ref{Fig-FW-LR}  of the main text. 

In Fig.~\ref{Fig-LT}, we study the long time value of $C(t)$, denoted by $C(\infty)$, where we calculate the average value of $C(t)$ (shown in Fig.~\ref{Fig-AAT} and \ref{Fig-AAT-LR})) over the last $50$ time steps. The results show good agreement with the theoretical scaling $\sim 1/L^m$ (Eq.~\eqref{eq-Ldm}), for both the tilted field and long-range Ising model.

To check the chaoticity of the models, we calculate the mean ratio $\langle r \rangle$ of the adjacent level
spacings 
\begin{equation}
\langle r\rangle=\frac{1}{N_{r}}\sum_{i=1}^{N_{r}}\frac{\min(\Delta_{i},\Delta_{i+1})}{\max(\Delta_{i},\Delta_{i+1})},
\end{equation}
where $\Delta_{i}=|E_{i+1}-E_{i}|$ denotes  the gap between two
adjacent eigenvalues $E_i$, and $N_r$ is the total number of gaps we consider. 

With periodic boundary conditions, both models we considered exhibit translation symmetry. The total Hilbert space can then be divided into different sectors with fixed quasimomentum $k$.
The sector $k=0$ and $k=\pi$ can be further split into subsectors using reflection symmetry, labeled by $p=\pm 1$. Furthermore, in the long-range model, due to the absence of longitudinal field, it has an additional $Z_2$ symmetry with respect to the flipping of the magnetic field in the z-direction.
As a result, each subsector can be even further split into subsubsectors, which can be labeled by $p_z = \pm 1$. In the computation of $\langle r \rangle$ (Fig.~\ref{Fig-Ar}), we consider the $(k=0,p=1)$ subsector for tilted field Ising model and the $(k=0,p=1,p_z=1)$ subsubsector for long-range Ising model. In both cases, we consider $1/2$ of the total eigenvalues in the middle of the spectrum. In both models, for all system sizes $L$ we consider, $\langle r \rangle$ is close to the Gaussian Orthogonal Ensemble (GOE) prediction $\langle r \rangle \approx 0.53$, indicating that the models are chaotic.

\end{appendix}

\end{document}